\pdfoutput=1
\documentclass[iop]{emulateapj}
\usepackage{apjfonts}
\usepackage{amssymb}
\usepackage{amsmath}
\usepackage{ctable}

\newcommand{\be}{\begin{equation}}
\newcommand{\ee}{\end{equation}}

\newcommand{\msun}{M_{\sun}}
\newcommand{\rhogas}{\rho_{\rm g}}
\newcommand{\grain}{_{\rm p}}
\newcommand{\rhograin}{\rho\grain}
\newcommand{\rhointernal}{\bar{\rho}_{\rm solid}}

\newcommand{\cs}{c_{s}}
\newcommand{\Rgrain}{a}
\newcommand{\tstop}{t_{\rm s}}

\newcommand{\Rcrit}{R_{\rm crit}}
\newcommand{\eddy}{_{e}}

\newcommand{\Veddy}{v\eddy}
\newcommand{\Teddy}{t\eddy}

\newcommand\plotonesize[2]
 {\centering \leavevmode \includegraphics[width={#2\columnwidth}]{#1}}

\shorttitle{Dust-to-Gas Fluctuations \&\ Metal Stars}
\shortauthors{Hopkins}
\slugcomment{Submitted to ApJ, July, 2014}
\begin{document}
\title{\vspace{-1.0cm}Some Stars are Totally Metal: A New Mechanism Driving Dust Across Star-Forming Clouds, and Consequences for Planets, Stars, and Galaxies}
%Some Stars are Totally Metal:\\ Driving Dust-to-Gas Variations in Star-Forming Clouds, and Consequences for Stellar Abundances, Carbon-Enhanced Stars, the Galactic UV Upturn, and Planet Formation\vspace{-0.5cm}}
\author{Philip F.\ Hopkins\altaffilmark{1}}
%\author[Hopkins]{
%\parbox[t]{\textwidth}{ 
%Philip F.~Hopkins\thanks{E-mail:phopkins@caltech.edu}\altaffilmark{1,2},
%\&\ Jessie L.~Christiansen\altaffilmark{3}} 
%\vspace*{6pt} \\
\altaffiltext{1}{TAPIR, Mailcode 350-17, California Institute of Technology, Pasadena, CA 91125, USA; E-mail:phopkins@caltech.edu} 

\label{firstpage}

\begin{abstract}
%\vspace{-0.2cm}

Dust grains in neutral gas behave as aerodynamic particles, so they can develop large local density fluctuations entirely independent of gas density fluctuations. Specifically, gas turbulence can drive order-of-magnitude ``resonant'' fluctuations in the dust density on scales where the gas stopping/drag timescale is comparable to the turbulent eddy turnover time. Here we show that for large grains (size $\gtrsim0.1\,\mu{\rm m}$, containing most grain mass) in sufficiently large molecular clouds (radii $\gtrsim1-10\,$pc, masses $\gtrsim10^{4}\,\msun$), this scale becomes larger than the characteristic sizes of pre-stellar cores (the sonic length), so large fluctuations in the dust-to-gas ratio are imprinted on cores. As a result, star clusters and protostellar disks formed in large clouds should exhibit significant abundance spreads in the elements preferentially found in large grains ({\small C, O}). This naturally predicts populations of carbon-enhanced stars, certain highly unusual stellar populations observed in nearby open clusters, and may explain the ``UV upturn'' in early-type galaxies. It will also dramatically change planet formation in the resulting protostellar disks, by preferentially ``seeding'' disks with an enhancement in large carbonaceous or silicate grains. The relevant threshold for this behavior scales simply with cloud densities and temperatures, making straightforward predictions for clusters in starbursts and high-redshift galaxies. Because of the selective sorting by size, this process is not necessarily visible in extinction mapping. We also predict the shape of the abundance distribution -- when these fluctuations occur, a small fraction of the cores may actually be seeded with abundances $Z\sim100\,\langle Z\rangle$ such that they are almost ``totally metal'' ($Z\sim1$)! Assuming the cores collapse, these totally metal stars would be rare ($1$ in $\sim10^{4}$ in clusters where this occurs), but represent a fundamentally new stellar evolution channel. 

\end{abstract}

\keywords{
star formation: general --- protoplanetary discs --- galaxies: formation --- galaxies: evolution --- hydrodynamics --- instabilities --- turbulence --- cosmology: theory
%\vspace{-1.0cm}
}

%\vspace{-1.1cm}
\section{Introduction}
\label{sec:intro}

For $\sim50$ years, essentially all models of star formation and molecular clouds have ignored a ubiquitous physical process -- namely that massive dust grains naturally experience large density fluctuations in turbulent, neutral media (so-called ``turbulent concentration''). 

It has long been known that, in atomic/molecular gas, dust grains -- which contain a large fraction of the heavy elements in the interstellar medium (ISM) -- behave as aerodynamic particles. As such, below some characteristic size scale, they are effectively de-coupled from the gas, and can (in principle) clump or experience density fluctuations independent from gas density fluctuations. But this process has largely been ignored in most of astrophysics. Recently, though, much attention has been paid to the specific question of grain density fluctuations and resulting ``grain concentrations'' in proto-planetary disks. When stirred by turbulence, the number density of solid grains can fluctuate by multiple orders of magnitude, even when the gas is strictly incompressible! This has been seen in a wide variety of simulations of both idealized ``pure'' turbulence and astrophysical turbulence in e.g.\ proto-planetary disks, both super and sub-sonic turbulence, including or excluding the effects of grain collisions, and in non-magnetized and magnetically dominated media \citep[see e.g.][]{elperin:1996:grain.clustering.instability,bracco:1999.keplerian.largescale.grain.density.sims,cuzzi:2001.grain.concentration.chondrules,johansen:2007.streaming.instab.sims,
youdin:2007.turbulent.grain.stirring,youdin:2011.grain.secular.instabilities.in.turb.disks,
carballido:2008.large.grain.clustering.disk.sims,bai:2010.grain.streaming.sims.test,bai:2010.streaming.instability,bai:2010.grain.streaming.vs.diskparams,pan:2011.grain.clustering.midstokes.sims,pan:2013.grain.relative.velocity.calc,bai:2012.mri.saturation.turbulence}. And grain clumping occurs similarly regardless of whether turbulent motions are self-excited \citep[the ``streaming'' instability;][]{youdin.goodman:2005.streaming.instability.derivation}, or externally driven via global gravitational instabilities, the magneto-rotational instability, convection, or Kelvin-Helmholtz instabilities \citep{dittrich:2013.grain.clustering.mri.disk.sims,jalali:2013.streaming.instability.largescales,hendrix:2014.dust.kh.instability,zhu:2014.dust.power.spectra.mri.turbulent.disks,zhu:2014.non.ideal.mhd.vortex.traps}. 

Indeed, {\em some} decoupling of the density of large dust grains from gas and small grains has been observed. \citet{kruger:2001.ulysses.large.grain.and.pioneer.compilation.data,frisch:2003.small.large.grains.decoupled.near.solar.neighborhood}, and subsequently \citet{meisel:2002.radar.micron.dust.ism.particles.many.large.grains,altobelli:2006.helios.large.local.interstellar.grains,altobelli:2007.cassini.confirms.large.microns.sized.interstellar.dust.grains,poppe:2010.new.horizons.confirms.ulysses.large.dust.measurements} have observed (via direct detection of micro-meteors and satellite impacts from interstellar grains) that the solar system appears to lie within a substantial (orders-of-magnitude) overdensity of large (micron-sized) grains \citep[for a review, see][]{draine:2009.ism.vs.local.dust.must.be.concentration.effects}. Some of this may stem from solar magnetic effects \citep[although they do not appear to be correlated; see][]{altobelli:2005.galileo.large.grain.fluxes.stable.over.decades}; but it could also easily result from the processes above. Within nearby (relatively small) molecular clouds, \citet{thoraval:1997.sub.0pt04pc.no.cloud.extinction.fluct.but.are.on.larger.scales,thoraval:1999.small.scale.dust.to.gas.density.fluctuations} identified large fluctuations in the ratio of large grains to gas on scales $\sim 0.001-0.1\,$pc, and others have subsequently seen similar effects \citep{abergel:2002.size.segregation.effects.seen.in.orion.small.dust.abundances,flagey:2009.taurus.large.small.to.large.dust.abundance.variations,boogert:2013.low.column.threshold.for.ice.mantle.formation}. Across different regions in the ISM, large variations in the relative abundance of large grains have been inferred from variations in the shapes of extinction curves and emission/absorption features \citep[see e.g.][]{miville-deschenes:2002.large.fluct.in.small.grain.abundances,gordon:2003.large.variations.extinction.curves.in.lmc.smc.mw.sightlines,dobashi:2008.dust.gas.variations.observed.over.lmc.gmcs,paradis:2009.large.variation.across.lmc.in.dust.size.distrib}.

In the terrestrial turbulence literature, this process (``preferential concentration'' of aerodynamic particles) is well studied. Actual laboratory experiments and measurements of turbulent systems (such as particulates in smokestacks, beads in water jets, raindrop formation in clouds, dust or water droplets in wind-tunnels and airfoil tests, and many more) ubiquitously demonstrate that turbulent gas is unstable to the growth of very large-amplitude (up to factor $\gg 10^{4}$, vaguely log-normally distributed) inhomogeneities in the dust-to-gas ratio \citep{squires:1991.grain.concentration.experiments,fessler:1994.grain.concentration.experiments,rouson:2001.grain.concentration.experiment,falkovich:2004.intermittent.distrib.heavy.particles,gualtieri:2009.anisotropic.grain.clustering.experiments,monchaux:2010.grain.concentration.experiments.voronoi}. The same has been seen in direct numerical simulations of these systems as well as idealized ``turbulent boxes''  \citep{cuzzi:2001.grain.concentration.chondrules,yoshimoto:2007.grain.clustering.selfsimilar.inertial.range,hogan:2007.grain.clustering.cascade.model,bec:2009.caustics.intermittency.key.to.largegrain.clustering,pan:2011.grain.clustering.midstokes.sims,monchaux:2012.grain.concentration.experiment.review}. And considerable work has gone into developing the analytic theory of these fluctuations, which has elucidated the key driving physics and emphasized the universality of these processes in turbulence \citep[see e.g.][]{yoshimoto:2007.grain.clustering.selfsimilar.inertial.range,hogan:2007.grain.clustering.cascade.model,zaichik:2009.grain.clustering.theory.randomfield.review,bec:2009.caustics.intermittency.key.to.largegrain.clustering}.

Here, we argue that the same physics should apply in star-forming molecular clouds, and argue that grain density fluctuations can occur at interesting levels in large clouds, with potentially radical implications for star formation and stellar evolution, as well as planet formation in the disks surrounding those stars.

\begin{figure}
    \centering
    \plotonesize{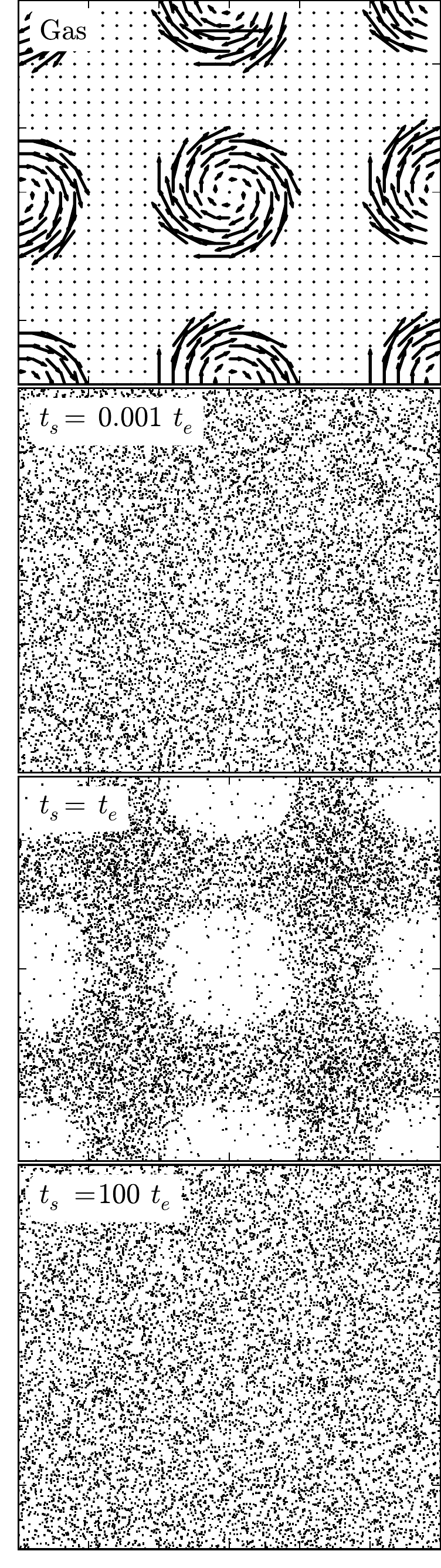}{0.67}
    %\vspace{-0.2cm}
    \caption{Illustration of the physics that drive dust-to-gas fluctuations in turbulence. We initialize $10^{4}$ dust particles that move aerodynamically (Eq.~\ref{eqn:stokes}), with random initial positions and velocities, and evolve them in a field of identical eddies for exactly one turnover time $\Teddy$. 
    {\em Top:} Gas velocity field. We consider a periodic box with identical vortices (each with circulation velocity $=r/\Teddy$ out to size $r_{0}$, and spaced in a grid, with no velocity in-between). The gas is in pure circulation so $\nabla\cdot {\bf v}=0$ everywhere (the gas density does not vary). 
    {\em Bottom:} Positions of the particles after $\Teddy$, for different stopping times $\tstop=(0.001,\,1,\,100)\,\Teddy$. If $\tstop\ll \Teddy$, dust is tightly-coupled to the gas (it circulates but cannot develop density fluctuations). If $\tstop\gg \Teddy$, dust does not ``feel'' the eddies, so moves collisionlessly. If $\tstop\sim \Teddy$, dust is expelled from vortices and trapped in the interstices.
    \label{fig:dust.vortex.demo}}
\end{figure}

%\vspace{-0.5cm}
\section{The Physics}

\subsection{Dust as Aerodynamic Particles}

In a neutral medium, dust grains couple to gas via collisions with individual atoms/molecules (the long-range electromagnetic or gravitational forces from individual molecules are negligible). In the limit where the grain mass is much larger than the mass of an individual atom (grain size $\gg$\,\AA), the stochastic nature of individual collisions is averaged out on an extremely short timescale, so the grain behaves as an aerodynamic particle. The equation of motion for an individual dust grain is the Stokes equation: 
\begin{align}
\label{eqn:stokes}\frac{{\rm D}{\bf v}}{{\rm D}t} &= -\frac{{\bf v}-{\bf u}({\bf r})}{\tstop}  +  \frac{1}{m_{\rm grain}}\,{\bf F}_{\rm ext}({\bf r})\\ 
\tstop &\equiv \frac{\rhointernal\,\Rgrain}{\rhogas\,\cs}
\end{align}
where ${\rm D}/{\rm D}t$ denotes the Lagrangian (comoving) derivative, ${\bf v}$ is the grain velocity, ${\bf u}({\bf r})$ is the local gas velocity, $\rhointernal\approx 2\,{\rm g\,cm^{-3}}$ is the {\em internal} grain density \citep{weingartner:2001.dust.size.distrib}, $\Rgrain = \Rgrain_{\mu}\,\mu{\rm m} \sim 10^{-4}-10$ is the grain size, $\rhogas$ is the mean gas density, $\cs$ the gas sound speed,\footnote{The expression for $\tstop$ is for the Epstein limit. It is modified when $\Rgrain\gtrsim 9\lambda/4$, where $\lambda$ is the mean free path for molecular collisions, but this is never relevant for the parameter space we consider.} and ${\bf F}_{\rm ext}$ represents the long-range forces acting on the grains (e.g.\ gravity).  It is important to note that most of the {\em mass} in dust, hence a large fraction of the total metal mass, is in the largest grains with $\Rgrain_{\mu}\sim0.1-10$ \citep[see][and references therein]{draine:2003.dust.review}.\footnote{This is true for both carbonaceous and silicate grains in detailed models such as those in \citet{lidraine:2001.dust.model.update,draine:2007.pah.model.update}, but also follows trivially from the ``canonical'' power-law grain size spectrum ${\rm d}N/{\rm d}a\propto a^{-(3.0-3.5)}$, giving ${\rm d}m/{\rm d}\ln{a} \propto a^{4}\,{\rm d}N/{\rm d}a$ \citep{mathis:1977.grain.sizes}. We should note that the sizes of the large grains containing most of the mass are still uncertain, and there are suggestions that these may be larger in dense, star forming regions. \citet{goldsmith:2008.taurus.gmc.mapping,schnee:2014.mm.sized.grains.in.star.forming.regions} see evidence of mm-sized grains in cold, local star-forming regions of Taurus and Orion, and \citet{grun:1993.interstellar.grains.large.sizes.collected,landgraf:2000.very.large.grains.from.interstellar.seen.by.spacecraft} have shown similar, interstellar grains are collected by spacecraft in the solar system. Smaller, but still $\gtrsim$ micron-sized grains are also implied by observations of X-ray halos \citep{witt:2001.xr.halos.imply.large.dust.grains}, extinction observations of local group star-forming regions \citep{de-marchi:2014.30.dor.large.micron.sized.grains.needed}, and other spacecraft collection/impact missions \citep{kruger:2001.ulysses.large.grain.and.pioneer.compilation.data,kruger:2006.updated.ulysses.dust.data,meisel:2002.radar.micron.dust.ism.particles.many.large.grains,altobelli:2005.galileo.large.grain.fluxes.stable.over.decades,altobelli:2006.helios.large.local.interstellar.grains,altobelli:2007.cassini.confirms.large.microns.sized.interstellar.dust.grains,poppe:2010.new.horizons.confirms.ulysses.large.dust.measurements}; such large grains have also been seen in SNe ejecta \citep{gall:2014.several.micron.grains.forming.directly.in.sne}. In any case, larger grains will amplify the effects we argue for here; but since this is uncertain, we will leave this maximum grain size as a free parameter in our model.}

Here $\tstop$ is the ``stopping time'' or ``friction time'' of the grain -- it comes from the mean effect of many individual collisions with molecules in the limit $m_{\rm grain}\gg m_{p}$ (for a $\Rgrain_\mu\sim1$, $m_{\rm grain} \sim 10^{13}\,m_{p}$).\footnote{Note that the grain-grain collision time is always larger than $\tstop$ by a factor $\gtrsim 1/Z\grain$, where $Z\grain$ is the grain mass fraction, so it is {\em not} what determines the grain dynamics.} On timescales $\ll \tstop$, i.e.\ spatial scales $\ll \tstop\,|{\bf v}|$, grains behave like collisionless particles (they do not ``feel'' pressure forces). It is only over timescales $\gg \tstop$ (spatial scales $\gg \tstop\,|{\bf v}|$) that the grains can be treated as tightly-coupled to the gas (e.g.\ as a fluid). Previous authors have pointed out that this decoupling should occur for massive grains, and can explain observed velocity decoupling between dust and gas in cold, star-forming clouds \citep[see e.g.][]{falgarone:1995.grain.decoupling.cold.clouds}. Others, such as \citet{murray:2004.local.sources.of.large.grains.and.10pc.freestreaming.distances}, have invoked this free-streaming (perhaps from a local source, or perhaps induced by local vortices or solar neighborhood gas velocity structures) as an explanation for the anomalous over-abundance of large interstellar grains detected by spacecraft. However, what have not been considered yet are the associated dust-to-gas density fluctuations which should occur in this regime.

%\vspace{-0.5cm}
\subsection{How Dust-to-Gas Fluctuations Occur}

Because of their imperfect coupling, grains can in principle fluctuate in density {\em relative to the gas}. This is a very well-studied problem with many review papers summarizing the physics \citep[see e.g.][and references therein]{zaichik:2009.grain.clustering.theory.randomfield.review,olla:2010.grain.preferential.concentration.randomfield.notes,gustavsson:2012.grain.clustering.randomflow.lowstokes}, so we will not present a complete derivation of these fluctuations here, but we will briefly summarize the important processes.

We illustrate this with a simple test problem in Fig.~\ref{fig:dust.vortex.demo}. Consider a turbulent eddy -- for simplicity, imagine a vortex in pure circulation ${\bf u}=u\eddy\,\hat{\phi} = (r\eddy/\Teddy)\,\hat{\phi}$, where $u\eddy$, $r\eddy$, $\Teddy = r\eddy/|u\eddy|$ are the characteristic eddy velocity, size, and ``turnover time'' (more exactly, the inverse of the local vorticity). Since by assumption we're taking the pure circulation case, the eddy produces zero change to the {\em gas} density. In Fig.~\ref{fig:dust.vortex.demo}, we initialize a simple grid of such eddies in two dimensions (though the results are essentially identical if we used three-dimensional Burgers vortices), and initialize a distribution of dust with randomly distributed initial positions and velocities (uniformly distributed), and evolve the system according to Eq.~\ref{eqn:stokes} for one $\Teddy$ (the characteristic lifetime of eddies). 

If $\tstop\gg\Teddy$, grains will simply pass through the eddy without being significantly perturbed by the local velocity field, so their density distribution is also unperturbed. If $\tstop\ll\Teddy$ is sufficiently small, the grains are efficiently dragged with the gas so cannot move very much {\em relative} to the gas in the eddy itself. But when $\tstop\sim\Teddy$, interesting effects occur. The grains are partly accelerated up to the eddy rotational velocity $v_\phi \sim u\eddy$. But this means they feel a centrifugal acceleration $\sim v_\phi^{2}/r$. For the gas, this is balanced by pressure gradients, but the grains feel this only indirectly (via the drag force); so they get ``flung outwards'' until the outward drag $\sim v_{\rm out} / \tstop$ balances centrifugal forces. Thus they ``drift'' out of the eddy at a terminal velocity $v_{\rm out}\sim \tstop \,v_\phi^{2}/r \sim (\tstop/\Teddy)\,u\eddy$. Since eddies live for order $\sim |\Teddy|$, the grains initially distributed inside $\sim r\eddy$ end up ``flung out'' to $\sim (1+\tstop/\Teddy)\,r\eddy$. When $\tstop\sim\Teddy$ this implies an order-unity median multiplicative change in the grain density {\em every time} the grains encounter a single turbulent eddy! 

Indeed, in laboratory experiments, numerical simulations, and analytic theory \citep[see references in][]{hopkins:2013.grain.clustering}, the full clustering statistics of dust grains in inertial-range turbulence depends only on the dimensionless parameter $\tstop/t\eddy(R)$, the ratio of the gas friction time $\tstop$ to the eddy time $t\eddy(R)$ as a function of scale $R$ (since, in full turbulence, eddies of all sizes exist, but they have different vorticities on different scales). In an external gravitational field or shearing disk, additional corrections appear but these are also scale free (depending on e.g.\ the ratio $\tstop/t_{\rm orbit}$). Generically, whenever $\tstop\sim t\eddy$, grains are partially accelerated to be ``flung out'' of regions of high vorticity by centrifugal forces, and preferentially collect in the interstitial regions of high strain \citep{yoshimoto:2007.grain.clustering.selfsimilar.inertial.range,bec:2008.markovian.grain.clustering.model,wilkinson:2010.randomfield.correlation.grains.weak,gustavsson:2012.grain.clustering.randomflow.lowstokes}. 

In a fully-developed turbulent cascade, there is a hierarchy of super-posed eddies or modes of various sizes; so the vorticity and strain fields are ``built up'' into a quasi-random field; in some regions they produce canceling (incoherent) effects, in others, coherent effects. Similarly, eddies survive about one turnover time, then dissipate, and new eddies form; so if we follow a Lagrangian group of grains we will see various encounters with eddies occur over a time-varying field. Crudely, this leads to a central-limit theorem-like behavior, where we build up a vaguely log-normal (because each encounter is {\em multiplicative}) dust-to-gas ratio distribution. Since a single eddy can produce changes of a factor of a few in the dust-to-gas ratio, it is easy to obtain order-of-magnitude fluctuations in the ``tails'' of this distribution \citep{hopkins:2013.grain.clustering}. Indeed, in both physical experiments and numerical simulations, factors of $>100-10^{4}$ fluctuations are common in terrestrial and protoplanetary disk turbulence \citep{bai:2010.streaming.instability,johansen:2012.grain.clustering.with.particle.collisions}. 

Many authors have pointed out that the relevant phenomena for this behavior are entirely scale free if the appropriate conditions are met \citep{cuzzi:2001.grain.concentration.chondrules,hogan:2007.grain.clustering.cascade.model,bec:2009.caustics.intermittency.key.to.largegrain.clustering,olla:2010.grain.preferential.concentration.randomfield.notes}. In \citet{hopkins:2013.grain.clustering} we derive the conditions for this to occur, as well as the width of the ``resonance'' region between timescales, and use this to build a model which reproduces the measured statistics of grain density fluctuations in both the simulations and laboratory experiments. Since at the very least this produces a good match to the simulation and experimental data, we will use it where needed for some more detailed calculations here.

%\vspace{-0.5cm}
\subsection{Generic PreConditions for Grain Density Fluctuations}\label{sec:preconditions}
Dust grains -- or any aerodynamic particles -- can undergo strong density fluctuations in gas provided three criteria are met: 

{\bf (1)}: The medium is predominantly atomic/molecular. Specifically, exchange of free electrons, Coulomb interactions, and/or coupling of ionized dust grains to magnetic fields will couple the dust-gas fluids more strongly than aerodynamic drag when the ionized fraction exceeds some threshold of order a percent (for conditions considered in this paper;\footnote{Note that we are explicitly interested in large grains here with sizes $\sim\mu{\rm m}$. PAHs and other very small grains (sizes $\sim$\AA) are photo-electrically coupled differently (and much more strongly) to the radiation field \citep[see e.g.][and references therein]{wolfire:1995.neutral.ism.phases}.} for a rigorous calculation see \citealt{elmegreen:1979.charged.grain.diffusion.gmcs}). At the densities and temperatures of interest here (molecular cloud cores), typical ionization fractions are $\lesssim 5\times10^{-8}$ \citep{guelin:1977.electron.abundance.gmc,langer:1978.gmc.electron.fracs,watson:1978.gmc.ionized.frac}; so this criterion is easily satisfied.\footnote{Provided this is satisfied, magnetic fields coupling to the {\em gas}, even in the super-critical regime, do not alter our conclusions. In fact \citet{dittrich:2013.grain.clustering.mri.disk.sims} show that Alfven waves and turbulence seeded by the fields in this limit actually {\em enhance} dust density fluctuations.} The case of (negatively) charged grain coupling to magnetic fields is more complicated, but for the (large) grains of interest here, we generally expect the Larmor period to be longer than the stopping time for equipartition magnetic fields, in which case we can neglect the magnetic acceleration in calculating the local response of grains to small-scale eddies;\footnote{For the conditions and grain sizes of interest here, grains are expected to vary between neutral and charge $Z_{\rm grain}\sim -1$ for an assumed ``sticking factor'' $s=1$ (probability that grain-electron collision leads to capture); but $s$ and other details are highly uncertain \citep{draine:1987.grain.charging}. If we assume that all grains spend most of their time charged, the relative importance of magnetic fields is given by the ratio of $\tstop$ to the Larmor period $t_{L} = 2\pi\,m_{\rm grain}\,c/(Z_{\rm grain}\,e\,|B|)$, which is $\tstop/t_{L} \approx 10^{-4}\,B_{\mu G}\,Z_{\rm grain}\,a_{\mu}^{-2}\,n_{100}^{-1}\,\delta v_{{\rm km\,s^{-1}}}$ (where $B_{\mu G}=|B|/\mu G$, $n_{100} = \langle n_{\rm gas}\rangle / 100\,{\rm cm^{-3}}$, and $\delta v_{{\rm km\,s^{-1}}}$ is the typical dust-gas relative velocity (between $\sim \cs$ and $\sim ({\cs^{2}+v_{\rm turb}^{2}})^{1/2}$, depending on scale). We can also write this as $\tstop/t_{L}\sim 0.01\,Z_{\rm grain}\,n_{100}^{-1/2}\,a_{\mu}^{-2}\,(|B|/|B|_{\rm eq})^{1/2}$ where $|B|/B_{\rm eq}=v_{A}^{2}/\delta v^{2}$ is the ratio of $|B|$ to an equipartition value. So we generally expect $\tstop \ll t_{L}$ even for the case where all grains are (weakly) charged. However, within plausible parameter space, the two effects can be comparable; in this limit the instabilities we describe have not been well studied.} And in fact, as emphasized by \citet{ciolek:1995.core.contraction.with.coupled.charged.dust}, grain-magnetic field coupling in a core contracting via ambipolar diffusion will tend to segregate out the small-size grains, changing the grain size distribution in a way that will increase the fluctuations we predict. We should also note that under the right conditions, direct coupling between the grains and magnetic fields, or photo-electric coupling of grains and radiation, can actually {\em enhance} and generate new instabilities driving grain density fluctuations \citep{lyra:2013.photo.electric.dust.instability}. But the behavior in these regimes is poorly understood, and needs further study.

{\bf (2)}: The medium is turbulent, with non-zero vorticity. There is no question that the ISM is turbulent under conditions we consider (with large Reynolds numbers). Within turbulence, the vorticity field (the solenoidal or $\nabla\times {\bf v}$ component) of the gas is what drives large grain density fluctuations independent of gas density fluctuations. The governing equations for grain density fluctuations over the inertial range \citep[see][]{zaichik:2009.grain.clustering.theory.randomfield.review} are independent of both whether or not the turbulence is super-sonic or sub-sonic, and whether the gas is compressible, except insofar as these change the ratio of solenoidal to compressive motions (essentially just re-normalizing the portion of the turbulent power spectrum that is of interest). In highly sub-sonic, incompressible flows, all of the turbulent power is in solenoidal motions; but even in highly compressible, isothermal, strongly super-sonic turbulence {\em driven} by purely compressive motions, the cascade (below the driving scale) quickly equilibrates with $\approx2/3$ of the power in solenoidal motions (this arises purely from geometric considerations; see \citealt{federrath:2008.density.pdf.vs.forcingtype,konstantin:mach.compressive.relation}). Thus up to a normalization of $2/3$ in the driving-scale rms velocity field, this is easily satisfied.

{\bf (3)}: There is a resonance between the grain stopping time $\tstop$ and turbulent eddy turnover time $t\eddy$: $\tstop\sim t\eddy$. We discuss this below.

\begin{figure}
    \centering
    \plotonesize{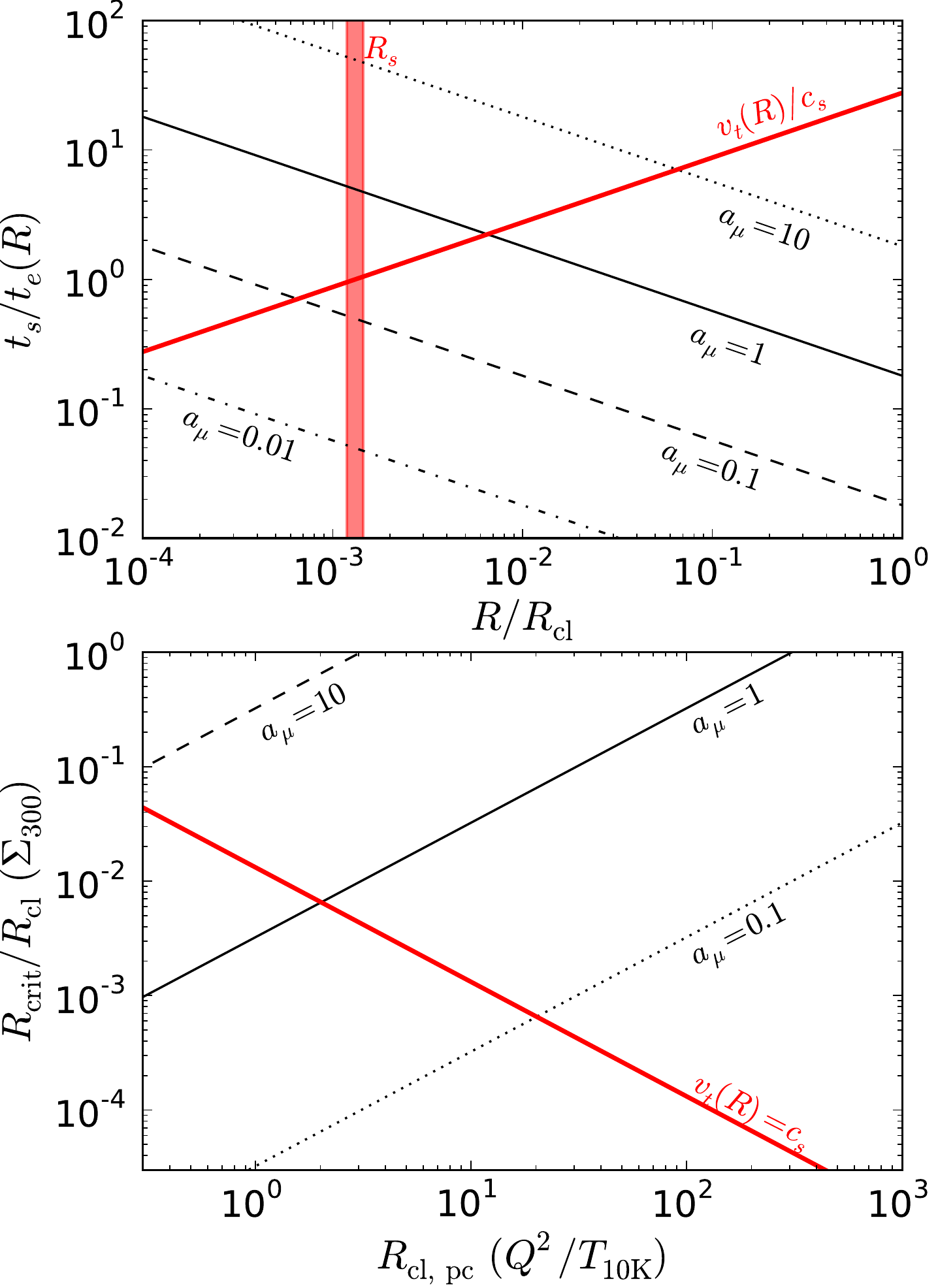}{0.99}
    %\vspace{-0.2cm}
    \caption{Conditions for strong dust-to-gas fluctuations in molecular clouds. 
    {\em Top:} Ratio of grain stopping time $\tstop$ to eddy turnover time $\Teddy$ for eddies of size $R$, relative to the cloud size $R_{\rm cl}$ (for a cloud with $Q=1$, $\Sigma_{\rm cl}=300\,\msun\,{\rm pc}^{-2}$, $T_{\rm min}=10\,$K, and $R_{\rm cl}=10\,$pc). Different curves assume different grain sizes $\Rgrain_\mu$. Thick red curve shows the ratio of the rms turbulent eddy velocity $v_{t}(R)$ to the sound speed; the scale where this $=1$ (the sonic length $R_{s}$) is the characteristic scale of cores. Small grains have $\tstop\ll \Teddy(R)$ for all $R \gtrsim R_{s}$; but large grains cluster on larger scales.
    {\em Bottom:} Scale $R_{\rm crit}$ where $\tstop=\Teddy(R)$ (normalized to the cloud size), for different grain sizes (lines as labeled), as a function of the cloud size in pc. Red line shows the sonic length in clouds of the same size: where the black lines exceed the red line, core-to-core dust-to-gas ratio variations are expected.
    \label{fig:tstop.vs.teddy}}
\end{figure}

\begin{figure}
    \centering
    \plotonesize{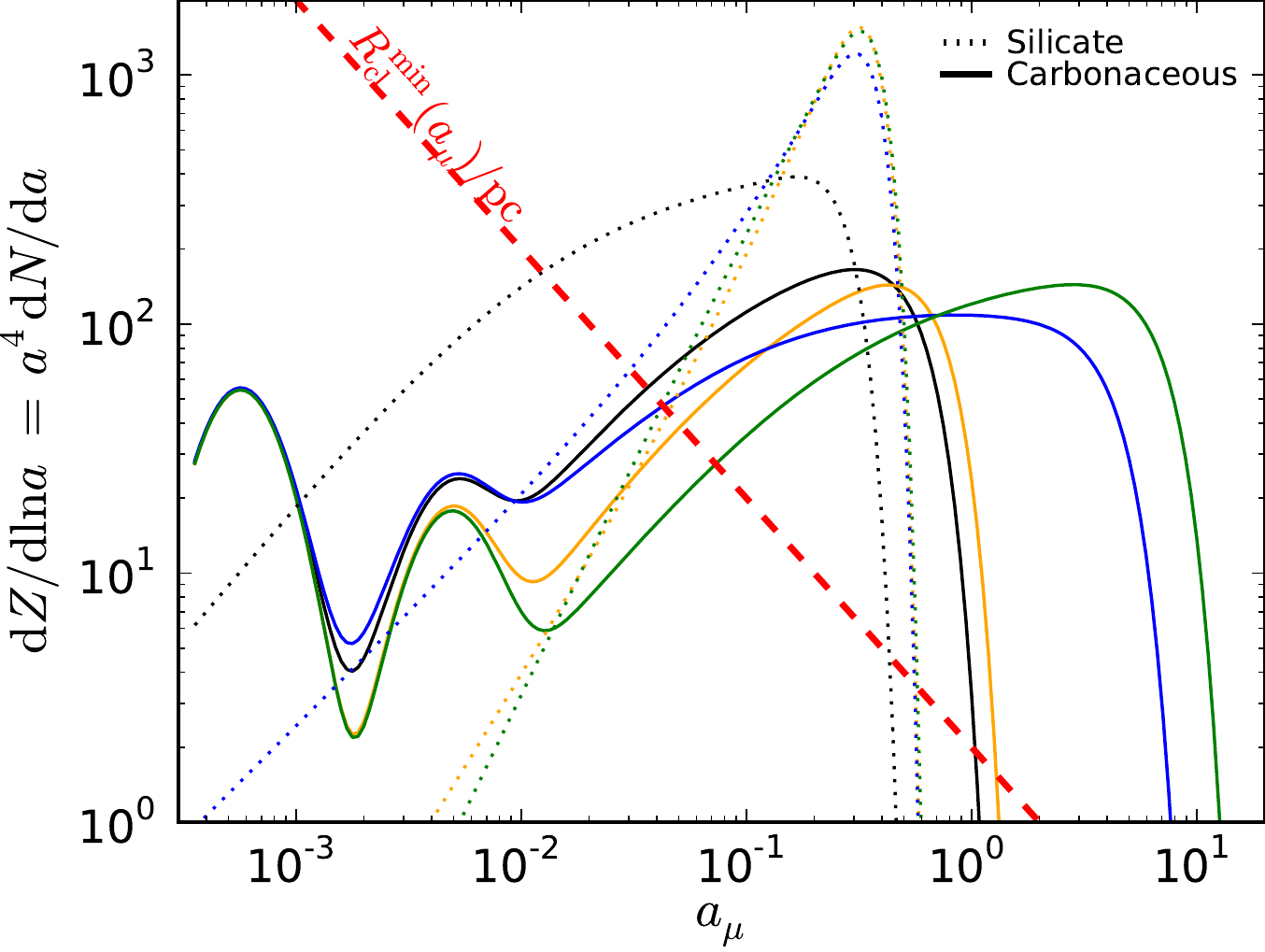}{0.99}
    %\vspace{-0.2cm}
    \caption{Size distribution of grains affected by turbulent clustering. We show the distribution of grain mass per log-interval in grain size $\Rgrain_\mu$ for carbonaceous (solid) and silicate (dotted) grains in several of the best-fit \citet{weingartner:2001.dust.size.distrib} models (the normalization is arbitrary, what matters here is the relative distribution in each curve). Different fits (or fits to different MW regions) give quite different results for the abundance of large grains, emphasizing their uncertain chemistry, but in all cases most of the mass is in large grains with sizes $\Rgrain_\mu\sim0.1-10$. Red dashed line shows the critical cloud size $R_{\rm crit}(\Rgrain_\mu)$ (above which grain density fluctuations are on super-core scales), for comparison. For a given cloud size, grains to the right of the corresponding $\Rgrain_\mu$ on the plot will experience core-to-core density fluctuations.
    \label{fig:grain.sizes}}
\end{figure}

%\vspace{-0.5cm}
\subsection{Specific Requirements in Molecular Clouds}

Provided condition {\bf (1)} is met, a grain obeys the Stokes equations (Eq.~\ref{eqn:stokes}). Now consider a molecular cloud which is super-sonically turbulent with mass $M_{\rm cl}$, size $R_{\rm cl}=10\,R_{10}$\,pc, surface density $\Sigma_{\rm cl}= M_{\rm cl}/(\pi\,R_{\rm cl}^{2})=\Sigma_{300}\,300\,{\msun}\,{\rm pc}^{-2} $, and sound speed $c_{s}\approx0.2\,T_{10}^{1/2}\,{\rm km\,s^{-1}}$ ($T_{10}=T/10\,K$ is the cold gas temperature, and we take a mean molecular weight $\approx2\,m_{p}$), and cloud-scale rms turbulent velocity $v_{\rm cl} = \langle v_{t}^{2}(R_{\rm cl}) \rangle^{1/2}$. For a constant-mean density spherical cloud, the Toomre $Q$ parameter is $Q\approx (2/\sqrt{3})\,v_{\rm cl}\,\Omega/(\pi\,G\,\Sigma_{\rm cl})$ with $\Omega=\sqrt{G\,M_{\rm cl}/R_{\rm cl}^{3}}$, and we expect $Q\sim1$ (even for a cloud in free-fall, this applies to within a factor $\sim2$). In a turbulent cascade, the rms velocity on a given scale follows $\langle v_{t}^{2}(R) \rangle^{1/2}\propto R^{p}$ with $p\approx1/2$ expected for super-sonic turbulence and observed in the linewidth-size relation \citep{burgers1973turbulence,bolatto:2008.gmc.properties,heyer:2009.gmc.trends.w.surface.density}. So the rms eddy turnover time on a given scale is $t\eddy \equiv R/\langle v_{t}^{2}(R)\rangle^{1/2} = (R_{\rm cl}/v_{\rm cl})\,(R/R_{\rm cl})^{1/2}$. 

We then have
\begin{align}
{\Bigl\langle}\frac{\tstop}{t\eddy(R)}{\Bigr\rangle} &= {\sqrt{3}\pi\,Q}\,\frac{G\,\rhointernal\,\Rgrain}{c_{s}\,\Omega}\,{\Bigl(}\frac{R}{R_{\rm cl}} {\Bigr)}^{-1/2}\\ 
&\approx 0.18\,Q\,\Rgrain_{\mu}\,{\Bigl(} \frac{R_{10}}{\Sigma_{300}\,T_{10}} {\Bigr)}^{1/2}\,{\Bigl(} \frac{R}{R_{\rm cl}} {\Bigr)}^{-1/2}
\end{align}
We plot this in Fig.~\ref{fig:tstop.vs.teddy}, for different grain sizes (and clouds with the characteristic parameters above). We compare the Mach number versus scale ($\langle v_{t}^{2}(R)\rangle^{1/2}/c_{s}$), which we discuss below. 

Note that even at the cloud-scale ($R=R_{\rm cl}$), 
%$\tstop\sim0.08\,t\eddy(R_{\rm cl})=0.08\,\Omega^{-1}$ 
$\tstop\sim0.18\,t\eddy(R_{\rm cl})=0.18\,\Omega^{-1}$ 
is sufficient to produce significant grain density variations \citep[see][]{johansen:2007.streaming.instab.sims,bai:2010.grain.streaming.vs.diskparams,dittrich:2013.grain.clustering.mri.disk.sims}. However, the fluctuations are maximized on the scale $\Rcrit$ where $\langle \tstop \rangle=\langle t\eddy(\Rcrit) \rangle$, so we can solve for the characteristic scale 
\begin{align}
\Rcrit = R(\langle\tstop\rangle=\langle t\eddy\rangle) = 0.31\,{\rm pc}\,\frac{\Rgrain_{\mu}^{2}\,Q^{2}\,R_{10}^{2}}{\Sigma_{300}\,T_{10}} \label{eqn:rcrit.vs.cloudparameters}
\end{align}
In the lower panel of Fig.~\ref{fig:tstop.vs.teddy}, we show this critical scale versus cloud size, again for different grain sizes.

If this scale ($\Rcrit$) is much less than the characteristic scale of the regions which collapse into pre-stellar cores ($R_{\rm core}$), then many independent small fluctuations will ultimately collapse into the same object (or stellar multiple) and be mixed, ``averaging out'' and imprinting little net abundance fluctuation in the stars (although they may seed interesting metallicity variations in the proto-stellar disk). However if $\Rcrit\gtrsim R_{\rm core}$, then independent collapsing regions will have different grain densities. A wide range of both simulations and analytic calculations show that $R_{\rm core}$ is characteristically the sonic length $R_{s}$, the size scale below which the turbulence is sub-sonic ($\langle v_{t}^{2}(R<R_{s})\rangle < c_{s}^{2}$; see \citealt{klessen:2000.cluster.formation,klessen:2001.sf.cloud.pdf,bate:2005.imf.mass.jeansmass,krumholz.schmidt,hennebelle:2008.imf.presschechter,hopkins:excursion.imf.variation,hopkins:excursion.imf,hopkins:excursion.clustering}). We show this scale for comparison in Fig.~\ref{fig:tstop.vs.teddy}. Below this scale, turbulence can no longer drive significant density fluctuations, so collapse proceeds semi-coherently (as opposed to a turbulent fragmentation cascade). For the cascade above, $R_{s} = R_{\rm cl}\,(c_{s}/v_{\rm cl})^{2}$, so 
\begin{align}
\frac{R_{\rm crit}}{R_{s}} = 23.5\,{\Bigl(}\frac{\Rgrain_{\mu}\,R_{10}\,Q^{2}}{T_{10}}{\Bigr)}^{2}
\end{align}
Seeding large fluctuations in cores ($\Rcrit>R_{s}$) therefore requires
\begin{align}
R_{\rm cl} &\gtrsim \frac{2}{3\pi\,Q^{2}}\,\frac{\cs^{2}}{G\,\rhointernal\,\Rgrain}\\
R_{10} &\gtrsim 0.2\,\Rgrain_{\mu}^{-1}\,Q^{-2}\,T_{10}\label{eqn:R10}
\end{align}
or (equivalently) 
\begin{align}
M_{\rm cl} \gtrsim 0.4\,\Rgrain_{\mu}^{-2}\times10^{4}\,\msun\,\Sigma_{300}\,T_{10}^{2}\,Q^{-4}
\end{align}
Note this is a cloud mass, not a cluster mass! The model here does not necessarily predict a cluster stellar mass threshold (which depends on the star formation efficiency), but a progenitor cloud mass threshold.

%\vspace{-0.5cm}
\subsection{Resulting Dust-to-Gas Density Fluctuations}
\label{sec:fluct.pdf}

Under these conditions, the density of grains averaged on scales $\sim \Rcrit$ will undergo large turbulent concentration fluctuations (fluctuations driven by local centrifugal forces in non-zero vorticity). Since this depends on the {\em vorticity} field, which is the incompressible/solenoidal component of the turbulence, this component of the fluctuations is entirely independent of gas density fluctuations (driven by the compressive field components).\footnote{There will, of course, be a component of the grain density fluctuations which traces advection through the compressive component of the turbulent field. This is not interesting for our purposes here, however, since the gas does the same so it imprints no variation in the gas-to-dust ratio.} Thus we expect large fluctuations in the dust-to-gas mass ratio $Z\grain\equiv \rhograin/\rhogas$. Specifically, only grains with sufficiently large $\Rgrain_\mu$ such that the condition in Eq.~\ref{eqn:R10} is satisfied are inhomogeneous on the relevant scales. In Fig.~\ref{fig:grain.sizes}, we plot the distribution of grain mass as a function of grain size $\Rgrain_\mu$, from various fits of observations presented in \citet{weingartner:2001.dust.size.distrib}; for comparison we show the critical cloud size $R_{\rm cl}$ where large core-to-core variations are expected at each $\Rgrain_\mu$. Although there are significant differences between different model fits, and different types of grains (silicate vs.\ carbonaceous), clearly grains with $\Rgrain_\mu\sim1$ contain most of the dust mass, and, correspondingly, an order-unity fraction of the total metal mass, so large $Z\grain$ fluctuations translate directly to large $Z$ variations in the appropriate species.

\citet{hopkins:2013.grain.clustering} show that, for systems with this range of $\tstop\,\Omega$ and $\tstop/t\eddy$, the {volume-weighted} variance in the logarithm of the dust-to-gas density $\ln{Z\grain}$ is given by\footnote{Here $\varpi(R/\Rcrit)$ is the ``response function'' -- the logarithmic density change per eddy turnover time induced by encounters between grains and an eddy with some $t\eddy$ -- in general it is a complicated function given in Table~1 therein, but it only depends on the dimensionless ratios $\tstop/t\eddy(R)$ and $\tstop\,\Omega$, and scales simply as $\approx (\tstop/t\eddy)$ when $\tstop\ll t\eddy$ and $\approx 1/\sqrt{2\,(\tstop/t\eddy)}$ when $\tstop\gg t\eddy$ (with a broad peak where $\varpi\approx0.3$ around $\tstop\approx t\eddy$).} 
\begin{align}
\label{eqn:S.zgrain} S_{\ln{Z\grain}} = \int_{R_{\rm min}\sim R_{s}}^{R_{\rm max}\sim R_{\rm cl}}\,2\,|2\,\varpi(R/\Rcrit)|^{2}\,{\rm d}\ln{R} \\ 
\sim 4.9\,-8\,{\Bigl(}\frac{\Rcrit}{R_{\rm cl}}-\sqrt{\frac{R_{s}}{\Rcrit}}{\Bigr)}
\end{align}
where the latter assumes $R_{\rm cl}\gg \Rcrit \gg R_{s}$; so for typical parameters the $1\,\sigma$ dispersion in metallicities is $\sim1\,$dex. Whether or not one believes this particular model, the numbers can be directly compared to simulations of idealized (hence essentially scale-free) ``shearing boxes'' of turbulence in \citet{johansen:2007.streaming.instab.sims} and \citet{dittrich:2013.grain.clustering.mri.disk.sims} which have $\tstop\,\Omega\sim0.1$ and otherwise similar (dimensionless) parameters to those calculated here; the simulations predict a very similar $\sim1\,$dex variance at the relevant scale.\footnote{For the same $\tstop/t\eddy$ and $\tstop\,\Omega$ simulated in \citet{johansen:2007.streaming.instab.sims,bai:2010.grain.streaming.sims.test}, the only parameter which affects the dust density distributions simulated, or enters the analytic calculation of those density distributions in \citet{hopkins:2013.grain.clustering}, and may differ between the proto-planetary disk problem (the problem which the simulations were originally used to study) and the molecular cloud case is the $\beta\propto1/\Pi$ parameter which depends on the cloud/disk-scale Mach number and gas pressure support. In the highly super-sonic limit, $\beta\rightarrow \infty$. But the effect of this in the simulations is simply to take to zero the small-scale ``cutoff'' below which the contribution of turbulent eddies to the grain clustering is damped; thus if anything the ``real'' clustering in the molecular cloud case should be slightly {\em larger} than in those simulations. We stress that all the numbers defining an absolute scale to the problem (e.g.\ gas density, dust grain size, orbital frequency) can be trivially re-scaled to the GMC problem here.}

Thus, the cloud is ``seeded'' with not just gas-density fluctuations (which form cores and determine where stars will form), but independent dust-to-gas ratio fluctuations, which are then ``trapped'' if they are associated with a collapsing core, so that the abundance in that core, and presumably the star formed, will be different. Note that if the conditions above are met, the characteristic timescale for the density fluctuations to disperse is always $\sim t\eddy(\Rcrit)$ (they cannot live much longer or shorter than the eddies generating them, before getting ``scattered'' into a new part of the distribution by new eddies). But, by definition, for a core to collapse, and overcome turbulent kinetic energy as well as shear and pressure terms to become a star, its collapse time {\em must} be shorter than $t\eddy$ (see \citealt{hennebelle:2008.imf.presschechter,padoan:2011.new.turb.collapse.sims,hopkins:frag.theory}), so the fluctuations are ``frozen into'' the cores.\footnote{It is also worth noting that the characteristic timescales (at the spatial scales of interest) for survival of density fluctuations and collapse of cores, and the timescale for ``seeding'' the density fluctuations ($\sim\tstop$) are much shorter than the timescale for e.g.\ grain drift/segregation by radiation pressure or grain formation/destruction, and shorter than the timescale for the ``parent'' cloud to be destroyed (a few $\Omega^{-1}$), so we can reasonably ignore those effects.}

Laboratory experiments \citep{monchaux:2010.grain.concentration.experiments.voronoi}, simulations \citep{hogan:1999.turb.concentration.sims,cuzzi:2001.grain.concentration.chondrules}, and the analytic models \citep{hopkins:2012.intermittent.turb.density.pdfs} show these dust density fluctuations have a characteristically log-Poisson shape: 
\begin{align}
P_{V}&(\ln{Z\grain})\,{\rm d}\ln{Z\grain} \approx \frac{\Delta N_{\rm int}^{m}\,\exp{(-\Delta N_{\rm int})}}{\Gamma(m+1)}\,\frac{{\rm d}\ln{Z\grain}}{\Delta} \\ 
\nonumber
m &= \Delta^{-1}\,{\Bigl\{}\Delta N_{\rm int}\,{\Bigl[}1 - \exp{(-\Delta)} {\Bigr]} - \ln{{\Bigl(} \frac{Z\grain}{\langle Z\grain \rangle}{\Bigr)}} {\Bigr\}}
\end{align}
where $\Delta N_{\rm int} \sim 2\,\langle\ln(R_{\rm cl}/R_{\rm crit})\rangle$ traces the dynamic range of scales which can contribute to grain density fluctuations, and $\Delta=\sqrt{S_{\ln{Z\grain}}/\Delta N_{\rm int}}$ is the rms weighted dispersion induced ``per eddy'' around $\sim\Rcrit$. In the limit where $N_{\rm int}\gg1$ (and to leading order around the mean $Z\grain$), this distribution is just log-normal. The mass-weighted distribution $P_{M}$ is just given by $P_{M}=Z\grain\,P_{V}$ (also approximately log-normal, with the same variance). 

However, although this distribution extends to $Z\grain\rightarrow0$, the actual metallicities $Z$ imprinted on the cores do not. The minimum $Z$ in a core will be given by the sum of the metals in a non-condensed phase (not in dust) and those in grains so small that their clustering scale is well below $R_{s}$. In the \citet{weingartner:2001.dust.size.distrib} models, the sum of non-condensed metals and grains with $\Rgrain_\mu<0.1$ is $1-f\grain\sim30-60\%$ of the total metal mass; so while large {\em positive} abundance enhancements may be possible, the largest decrements will be factor of a few. More generally, if for some species the mean mass fraction in large grains is $\langle Z\grain \rangle=f\grain\,\langle Z \rangle$, then the abundance of a given core will be $Z_{c}/Z = (1-f\grain) + f\grain\,(Z\grain/\langle Z\grain \rangle)$; so the variance in $Z$ imprinted on cores is reduced with $f\grain$. To first order, this is just
\begin{align}
S_{\ln{Z_{c}}} \approx f\grain^{2}\,S_{\ln{Z\grain}}
\end{align}

Also from \citet{hopkins:2013.grain.clustering}, the maximum fluctuation amplitude we expect is given by 
\begin{align}
\ln{{\Bigl(}\frac{Z\grain^{\rm max}}{\langle Z\grain\rangle}{\Bigr)}} &\approx 2\,
 \int_{R_{s}}^{R_{\rm cl}}\,[1-\exp{(-2\,\varpi)}]\,{\rm d}\ln{R}\\
 & \sim 9.7\,\exp{[-1.12\,(R_{s}/\Rcrit)^{1/4}]} \sim 4.5-5.5
\end{align}
where the latter assumes $R_{\rm cl}\gtrsim100\,\Rcrit$ (though this enters weakly). Thus we can obtain $Z\grain^{\rm max}\sim 100\,\langle Z\grain \rangle$. This also agrees well with simulations of the similar parameter space; and although it seems large, it is actually much smaller than the largest fluctuations which can be obtained under ``ideal'' circumstances ($\tstop\,\Omega\approx1$, where $Z\grain^{\rm max}\sim10^{4}\,\langle Z\grain\rangle$ is possible; see \citealt{bai:2010.streaming.instability,johansen:2012.grain.clustering.with.particle.collisions}). Comparing to the simulations or simply plugging in this value to the log-Poisson distribution above, we see that the {\em mass} fraction associated with such strong fluctuations is small, $\sim 10^{-4} - 10^{-3}$. However, these would be extremely interesting objects; if the seed $\langle Z\grain \rangle\sim Z_{\odot}\sim0.01-0.02$, then in these rare collapsing cores, $Z\grain\gtrsim1$ -- i.e.\ the collapsing core is {\em primarily} metals!

\begin{figure}
    \centering
    \plotonesize{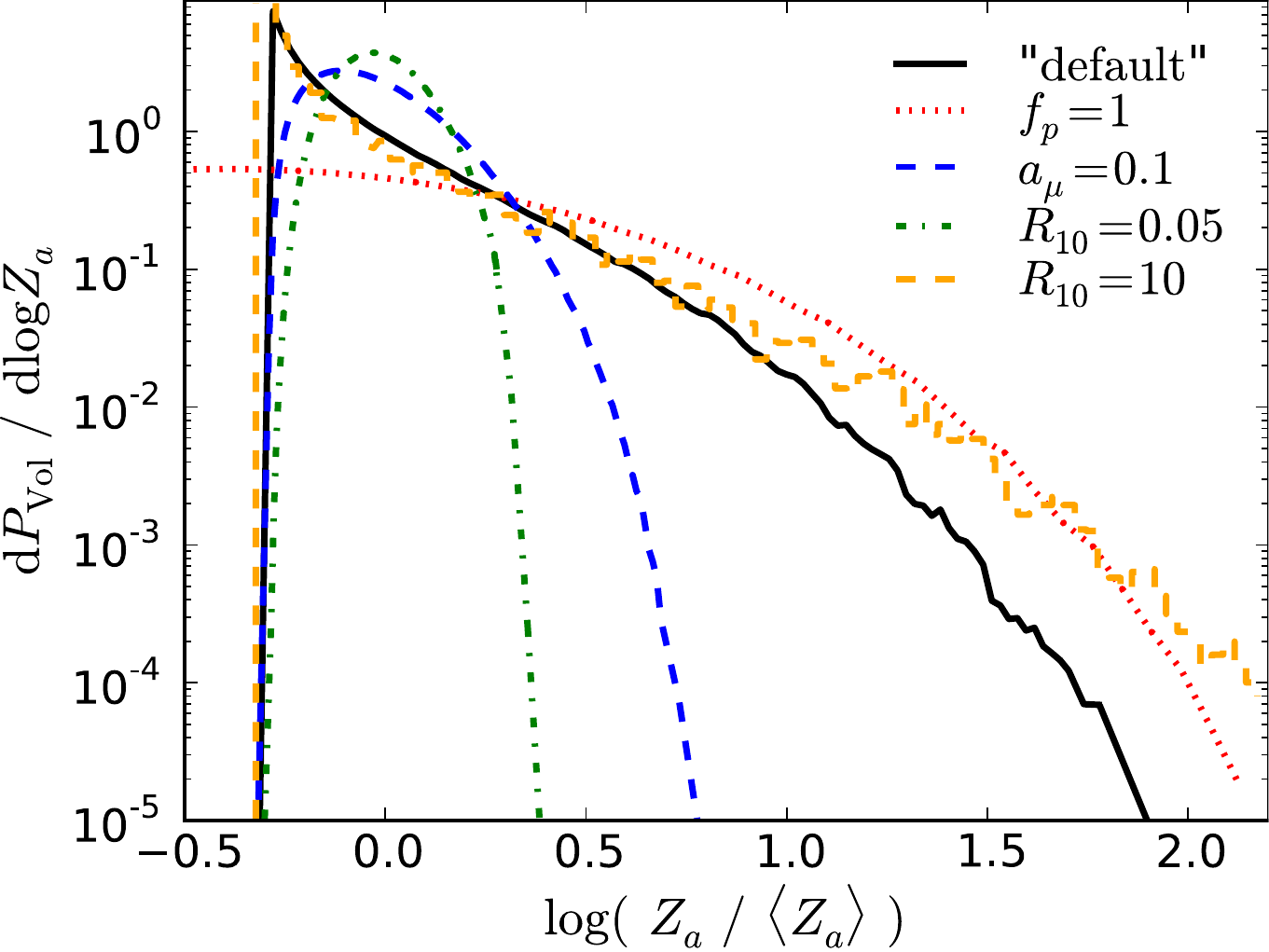}{0.99}
    %\vspace{-0.2cm}
    \caption{Predicted volumetric distribution of the metallicity fluctuations ($Z_{a}$, relative to the cloud-mean $\langle Z_{a}\rangle$) seeded by grain density fluctuations. In the ``default'' case, we assume that a fraction $f_{p}=0.5$ of the metal species of interest is condensed into grains with size $\ge a_{\mu}=1$, in a cloud with $R_{10}=T_{10}=\Sigma_{300}=1$. Since $1-f_{p}=1/2$ of the metallicity is in small grains and/or gas, it imposes a hard cutoff at low-$Z_{a}$, but positive fluctuations extend to large values $\sim 30\,\langle Z_{a} \rangle$. We then consider the same case but with all metals in large grains ($f_{p}=1$; red dotted); this extends the large-$Z_{a}$ by a factor $1/f_{p}$, and leads to a predicted low-$Z_{a}$ tail. We also compare a case identical to our ``default'' but with smaller grains $a_\mu=0.1$ (blue dashed) or a much smaller cloud ($R_{10}=0.05$; green dot-dashed); here the resulting fluctuations are much smaller (factor $<2$). Finally we consider the ``default'' case but with $R_{10}=10$ (a $100\,$pc cloud; orange dashed); now the largest positive fluctuations extend to factors of $\sim100$ -- the resulting cores would actually be {\em predominantly} metals by mass!
    \label{fig:z.distrib}}
\end{figure}

In Fig.~\ref{fig:z.distrib}, we calculate the full PDF of grain density fluctuations, integrating the contribution from the ``response function'' from the top scale $R_{\rm cl}$ down to the core scale $R_{s}$, according to the exact functions derived and fit to simulation data in \citet{hopkins:2013.grain.clustering}. This confirms our simple (approximate) expectations above. We assume a fraction $f_{p}$ of the metals are concentrated in grains, which (by assumption) imprints a cutoff in the negative metal density fluctuations at $1-f_{p}$. But the positive-fluctuation distribution is log-Poisson (approximately log-normal), and extends to larger values of $S_{\ln{Z\grain}}$ as the ratio of $R_{\rm crit}/R_{s}$ increases (i.e.\ with larger clouds and/or grain sizes). The largest positive fluctuations are predicted reach $Z\sim 30-300\,\langle Z \rangle$, depending on the detailed conditions, with a fractional probability by volume (i.e.\ probability that a random location has such a strong fluctuation) of $\sim 10^{-3}-10^{-4}$. 

Realistically, feedback effects from the back-reaction of dust grains pushing on gas become important in the limit where the mass density of grains exceeds that of gas (and this is not taking into account in the simple model here). Simulations show that these effects will tend to saturate the maximum dust density concentrations around $\rhogas\sim\rhograin$ or $Z\grain\sim1$ \citep[see e.g.][]{hogan:2007.grain.clustering.cascade.model,johansen:2007.streaming.instab.sims}. So we expect the most extreme fluctuations to be comparable in grain and gas mass (i.e.\ saturating at $\sim 1/2$ the mass in metals), and we should see a cutoff in Fig.~\ref{fig:z.distrib} at the high-$Z\grain$ end. But this is still very interesting for our purposes.

%\vspace{-0.5cm}
\section{Predictions}
\label{sec:predictions}

This makes a number of unique predictions: 

{\bf (1)} Stars formed in small clouds are not affected (chemically homogeneous). The key requirement is that the size of the progenitor GMC exceed $R_{\rm cl}\gtrsim 2\,{\rm pc}\,Q^{-2}\,(T_{\rm min}/10\,K)\,\Rgrain_{\mu}^{-1}$, where $\Rgrain_{\mu}$ is the grain size which contains most of the metal mass ($\sim0.1-1$, depending on the species). This corresponds to a GMC mass $\gtrsim 10^{4}\,\Rgrain_\mu^{-2}\,\msun$, for typical cloud densities in the local group.

{\bf (2)} Larger/more massive clouds can experience large dust-to-gas ratio (hence, abundance) fluctuations from core-to-core. The rms fluctuations could reach as large as $\sim1\,$dex under the right conditions. And these fluctuations should be ``frozen into'' the cores that collapse under self-gravity, meaning they will manifest in the stars formed from those cores. 

{\bf (3)} These dust density fluctuations will {\em not} obviously manifest as variation in the dust extinction. While most dust {mass} (or volume) is in large grains ($\Rgrain_\mu\sim0.1-1$), most of the surface area (hence contribution to extinction in most wavelengths) is in small grains ($\Rgrain_\mu < 10^{-3}$; the small ``bump'' in Fig.~\ref{fig:grain.sizes}). The small grains cluster only on scales $\sim10^{6}$ times smaller (sub-au), so trace the gas smoothly on core and GMC scales. Only dust features which are specifically sensitive to large grains will exhibit large spatial-scale fluctuations. Sub-mm observations (e.g.\ measurements of the spectral slope $\beta$), and other long-wavelength probes, may be sensitive to the features of larger dust, and so represent a way forward to constrain this process directly. Scattering, visible in e.g.\ X-ray halos and polarized light, also provides a direct constraint on the population of large grains; however this is more challenging in the dense (GMC-type) regions of interest here. Indirect (model-dependent) constraints, from combined constraints on depletion of heavy elements and extinction observations, are also possible \citep[e.g.][]{jenkins:2009.depletion.patterns.vs.densities.in.ism.grains}. While challenging, all the of these probes are possible, and we discuss the resulting constraints in \S~\ref{sec:dust.to.gas.in.gmcs}.

{\bf (4)} Because it is large grains that are preferentially clustered, certain metal species are preferentially affected. Inverting Eq.~\ref{eqn:R10}, we see that the minimum grain size which will be highly clustered on the relevant scales goes as $\Rgrain_\mu \sim 0.2\,T_{10}\,R_{10}^{-1}\sim 1.6\,(M_{\rm cl}/10^{4}\,\msun)^{-1/2}\,\Sigma_{300}^{-1/2}\,T_{10}$. 

Carbon and other light elements ({\small C, O}), concentrated in large composite grains and ices forming on the grains under GMC conditions ($\Rgrain_\mu\sim1-20$ in Fig.~\ref{fig:grain.sizes}; see e.g.\ \citealt{hoppe:2000.meteorite.dust.constraints}), should presumably exhibit the largest fluctuations, and these fluctuations will appear ``first'' in lower-mass clusters ($M_{\rm cl}\gtrsim10^{4}-10^{5}\,\msun$).\footnote{{\small N} may be an interesting case, since while it is probably concentrated in organic refractories in the largest grains \citep{zubko:2004.dust.chemistry.models}, it is mostly in gas-phase, so the variance may be smaller by a factor of $\sim4-8$ than in {\small C} or {\small O}.} Slightly heavier species ({\small Al}) will follow in clusters $\gtrsim 10^{5}\,\msun$. Heavier elements typical of Type-II SNe products and iron-peak elements ({\small Ca, Fe}) appear to be largely concentrated in small carbonates and oxides, with $\Rgrain_\mu\lesssim0.1$ \citep{tielens:1998.dust.is.amorphous.iron.poor.silicates,molster:2002.silicates.are.iron.poor,kemper:2004.silicates.are.amorphous.and.olivine.pyroxene} -- these abundances will only fluctuate in the most extreme clouds with masses $M_{\rm cl}\gtrsim10^{7}\,\msun$. {\small Si} ($\Rgrain_\mu\sim0.2-0.4$; see Fig.~\ref{fig:grain.sizes}) will be intermediate between these and {\small Al}.\footnote{For species in small grains, we can crudely approximate their size distribution following the added ``small grain'' component in \citet{draine:1998.spinning.dust}, and assume the minimum size of grains affected by these mechanisms scales as $\Rgrain_{\mu}\propto M_{\rm cl}^{-1/2}$ and is above the (nanometer-sized) ``peak'' in this lognormal-like distribution (above which ${\rm d}N/{\rm d}a_{i}$ for this sub-population falls approximately as $\propto a_{i}^{-4.5}$). If we insert these scalings into our models from Fig.~\ref{fig:z.distrib}, we then expect the fraction of the small-grain mass involved in fluctuations to increase with $M_{\rm cl}$, giving a predicted abundance variation $\delta [Z]_{i} \sim M_{\rm cl}^{0.1-0.2}$, i.e.\ the scatter increases by $\sim 0.04-0.08\,$dex per cluster absolute magnitude. This appears to consistent with observed {\small Fe} spreads in various cluster populations \citep[e.g.][]{carretta:2009.iron.spread.vs.gc.mass}.}

{\bf (5)} Very light elements such as {\small Li}, or chemically inert elements such as {\small He} and {\small Ne}, not being concentrated in grains, will not vary in abundance by this mechanism. In contrast, most models for abundance variation via self-enrichment (from mass loss products) or stellar evolution predict large {\small He} and {\small Li} variation. This gives a clear diagnostic to separate these processes. Some clusters certainly exhibit large {\small He} spreads \citep{piotto:2007.he.variations.gc.ngc2808}, but others show surprisingly small {\small He} and {\small Li} spreads, despite other abundance variations \citep{monaco:2012.lithium.variation.gc.constraints}.

{\bf (6)} The {\em shape} of the distribution of abundances seeded by turbulent concentration is approximately log-normal, or more accurately log-Poisson. This is consistent with some observed clusters, and may provide another way to separate variations owing to this process from self-enrichment processes which predict a more bi-model distribution \citep[see][]{carretta:2009.gc.abundance.var.compilation,carretta:2009.gcs.na.o.corr,carretta:2013.ngc362.gc.split.rgb}. The process seeding the variations (turbulent concentration) is also stochastic, which means that there is {intrinsic} cluster-to-cluster variation (even at the same initial mass and abundance).

{\bf (7)} Large negative abundance variations will not occur, because cores will still retain a minimum metallicity corresponding to the metals in non-condensed phases and very small grains. Thus the abundance distribution will be truncated at a minimum of a factor a couple below mean. 

{\bf (8)} At higher cloud surface densities ($\Sigma$) and/or minimum gas temperatures ($T$), the cloud mass threshold required for these variations increases. Both $\Sigma$ and $T$ do appear to increase with the overall gas density and star formation rate density -- for example in the nuclei of ultra-luminous infrared galaxies/starbursts and high-redshift galaxies. For typical ULIRG-like conditions, $Q\sim1$ remains constant, but $T_{10}\sim5-7$ and $\Sigma_{300}\sim10-30$ -- so this process requires $R_{10}\gtrsim1\,\Rgrain_\mu^{-1}$ or $M_{\rm cl}\gtrsim 10^{6}\,\Rgrain_\mu^{-2}\,\msun$. However, the characteristic Jeans length/mass also increases under these conditions (for a detailed comparison, see \citealt{hopkins:excursion.imf.variation}), up to several hundred pc (and $\gtrsim10^{8}\,\msun$); the thermal Jeans length/mass increases more rapidly with temperature then the critical cloud radius, for example. So whether older clusters exhibit weaker or stronger abundance variations will depend on the types of galaxies in which they formed, in detail, but we generically expect this process to be easier in the early Universe and extreme environments. 

{\bf (9)} Unlike abundance variation stemming from stellar evolution, binary effects, pollution, or second-generation star formation, the variations predicted here are imprinted {\em before} stars form and evolve, and so are independent of whether or not the final cluster remains bound. Thus we predict similar trends in open clusters (OCs) and associations (not just bound clusters). 

{\bf (10)} A small sub-population of proto-stellar cores (one in $\sim10^{4}$ within the clouds where this occurs) is predicted with metallicities $Z\grain\gtrsim100\,\langle Z\grain \rangle$. 
If the initial cloud metallicity is solar ($\langle Z\grain \rangle \gtrsim 0.01$), this implies $Z\grain\sim1$ might be reached -- cores would collapse into stars which are almost ``totally metal''! This fundamentally represents a novel stellar evolution channel, one which has not yet been well-explored.

As noted in \S~\ref{sec:fluct.pdf}, back-reaction of dust pushing on grains will saturate the dust concentrations at an order-unity fraction of the gas mass. Assuming a $50-50$ split between gas and dust, and solar abundances for the gas, this implies a maximum core metallicity in e.g.\ carbon of {\small [C/H]}$\sim 1.3$ (if grains also have solar abundances) or {\small [C/H]}$\sim 0.9$ (if grains have the mean abundances of ices in \citealt{draine:2003.dust.review}).

The detailed collapse process of such cores is not entirely clear, since it depends on the behavior of grains in the collapsing cloud. If the grains can stick efficiently, they will grow to large sizes, forming a self-gravitating cloud of pebbles. The cloud would thus go through an intermediate or adolescent ``rock'' phase before progressing to its more mature ``metal-star'' phase. More likely, the grains will shatter as the cloud collapses (since their collision velocities would be comparable to the gravitational free-fall velocity, hence large). This would reduce them to small grains and gas-phase metals, enhancing the cloud cooling, thus further promoting rapid collapse, until the cloud forms a star. More detailed calculations of this process, which argue that shattering should, in this limit, lead to efficient cooling and rapid collapse, are presented in \citet{chiaki:2014.critical.dust.abundance.for.cooling} and \citet{ji:2014.silicate.dust.cooling.for.metal.poor.star.criterion.and.tests},; details of the relevant chemical networks for a hyper metal-rich dust mixture trapped in a turbulent vortex are calculated in \citet{joulain:1998.local.chemistry.in.turb.vortices.special.for.molecules}. In all cases, these calculations agree that cooling and collapse should be efficient.

Once such a star forms, its properties would be quite interesting. While more work is clearly needed, we can gain some insight from idealized models. \citet{paczynski:1972.carbon.stars,paczynski:1973.pure.carbon.star.stability} calculated the main sequence for pure-carbon stars ($Z\grain\rightarrow\infty$). They argued that the most important difference between ``normal'' stars and these (simplified) models on the main sequence is the mean molecular weight ($\mu\approx 1.730$ and $\mu\approx 0.624$, respectively); for a (more likely) mixture with $Z\grain\sim1$ and solar-abundance gas and grains, $\mu\approx 1.27$, so the corresponding linear model if we assume the abundances are perfectly-mixed is closer to the Helium main sequence ($\mu=1.343$; \citealt{gabriel:1968.pure.carbon.star.stability,kozowski:1973.helium.carbon.stellar.mixes,kippenhahn:2012.stellar.evolution.book}). In either case, because of their enhanced $\mu$, the stars should be compact and (presumably) hot. Qualitatively, we find similar results based on preliminary calculations using the {\small MESA} code \citep{paxton:2013.mesa.code.update}. 

Assuming they carried or accreted some gaseous envelope, the massive ($\gtrsim2-3\,M_{\sun}$) versions of such metal stars would burn through their residual light elements quickly (much faster than their less-metal rich counterparts). Their luminosities ($\gtrsim 10^{3}\,L_{\sun}$) and effective temperatures ($\gtrsim 20,000\,$K) would be similar to O-stars; however in \S~\ref{sec:galaxy.abundance.var} we estimate that their relative abundance would be lower by a factor $\sim 10^{5}$. Given these luminosities, their main-sequence lifetimes are predicted to be quite short ($<10^{6}-10^{7}$\,yr), after which the intermediate-mass systems would transition smoothly to the C-O white dwarf sequence \citep{paczynski:1972.carbon.stars}. As a result they would not be recognizably different from other (slightly more evolved) massive stars which have already depleted their light elements. The most massive may explode as unusual SNe. Extremely-massive versions of such stairs may in fact be interesting pair-instability SNe candidates (since they would reach large {\small O} cores quickly, with a very different mass-loss history; A.~Gal-Yam, private communication). 

At dwarf masses, if we only consider the difference in mean molecular weight from low-metallicity stars, the resulting stars would be more compact and closer to the helium main sequence, with $T_{\rm eff}\sim 20,000-50,000\,$K and $L\sim 0.1-300\,L_{\sun}$ for $M\sim 0.3-1.5\,M_{\sun}$ (corresponding to main-sequence lifetimes of $\sim 10^{8}-10^{10}$\,yr, although these could be much shorter if neutrino cooling from the core is efficient; \citealt{paczynski:1972.carbon.stars}).\footnote{If the stars were truly ``pure metal,'' then any below $M\lesssim 0.8\,M_{\sun}$ would be unable to trigger burning of carbon or heavier elements, and collapse directly into a ``normal'' white dwarf (except perhaps for some unusual isotopic ratios in certain elements).} They would still begin their lives burning hydrogen, but this would proceed more rapidly if they are compact. However, this ignores the high opacity of the object, which means that the outer envelope could be inflated to large radii $\sim 1\,R_{\sun}$, giving much lower $T_{\rm eff}\sim 5000-30,000\,$K \citep{giannone:1968.two.component.stellar.envelope.models,kozowski:1973.helium.carbon.stellar.mixes}. Stars similar to this (metal-enhanced cores with relatively small light-element envelope masses) are in fact already known, and tend to produce ``extreme'' horizontal branch (EHB) or ``blue'' horizontal branch (BHB) stars; 
%as well as ``failed AGB'' or ``AGB-manqu{\'e}'' stars 
\citep{horch:1992.uv.excess.origins,liebert:1994.blue.hb.stars.ngc6791,landsman:1998.uv.upturn.relation.to.starclusters,greggio:1999.uv.excess.review,yong:2000.hot.subdwarf.star.origins}. With such high carbon abundances expected in their envelopes, the most metal-rich stars -- once they have quickly consumed their hydrogen reservoirs -- could also masquerade as R Cor Bor (RCB) or hydrogen-deficient carbon (HdC) stars (which commonly show {\small [C/Fe]}$\sim 1.5-2.5$, {\small [C/H]}$\gtrsim 5$, and have abundances of $\sim1$ per $\sim 10^{6}$ stars; \citealt{asplund:1998.rcorbor.abundances,pandey:2004.rcorbor.carbon.abundances,rao:2005.surface.abundance.rcorbor.review,kameswara-rao:2007.rcorbor.abundance.review,garcia-hernandez:2009.rcorbor.abundances,garcia-hernandez:2009.lmc.rcorbor.abundances,hema:2012.rcorbor.abundance.ratios}). The observed examples of these stars are often {assumed} to have typical total metal abundances with the residual mass in {\small He}, and while this is probably the case for many of these stars (as a result of their formation in white dwarf mergers), it is not actually measured, and successful models may be constructed with total $Z\sim 1$. 

Finally, we stress that this all only applies for the most extreme possible abundance fluctuations: more modest, but still highly metal-enhanced populations (with, say, $Z\sim 5-10\,Z_{\sun}$) -- the much more likely outcome -- would actually be cooler and dimmer compared to solar-metallicity stars of the same mass (they would have only slightly different mean molecular weights, but much higher atmospheric opacities).

%\vspace{-0.5cm}
\section{Speculations on the Relevance to Observed Stars, Planets and Galaxies: Does This Matter?}

In this section, we briefly outline some potential consequences of this mechanism for stars, galaxies, and planets, with a particular focus on some areas where the effects may be observable. However, we stress that this discussion will necessarily be highly speculative; better understanding of stellar evolution with unusual abundance patterns, and the chemistry of large grains, as well as improved observations with much larger samples, will be needed for any definitive conclusions.

%\vspace{-0.5cm}
\subsection{Carbon-Enhanced Stars}
\label{sec:carbon.stars}

The most natural result of the process described here is a population of stars enhanced specifically in the elements associated with the largest grains: {\small C} in particular (since these are carbonaceous grains; the largest often graphite chains). It has long been known that there are various populations of anomalously carbon-rich stars \citep[see e.g.][]{shane:1928.carbon.star.spectra}; more recently, large populations of carbon-enhanced metal-poor stars (CEMPS) have been identified with {\small [C/Fe]}$\sim 0-4$.

Among the ``normal metallicity'' carbon-rich population, many features follow naturally from the predictions here. They have high {\small [C/Fe]}, but also usually have {\small [C/O]}$>1$  -- this is highly non-trivial, since while {\small [C/O]}$>1$ is generically true inside a carbonaceous grain, it holds almost nowhere else in the Universe. Among the carbon-rich population, the ``early-R'' sub-population (which constitutes a large fraction of all carbon-rich stars; \citealt{blanco:1965.carbon.star.locations,stephenson:1973.s.star.catalogue,bergeat:2002.carbon.star.luminosity.function}) has properties which most naturally follow from this scenario: enhanced {\small C}, but no enhancement in s-process elements, little {\small Li} enhancement, and an absence of obvious companion or merger signatures \citep[][and references therein]{dominy:1984.carbon.star.chemistry,mcclure:1997.carbon.r.stars.no.binaries,knapp:2001.r.star.abs.mags,zamora:2009.carbon.r.star.chemistry}. Given the relatively modest {\small C} enhancements needed ($\sim0-1\,$dex), their abundance by number (a few to tens of percent of the carbon-star population), locations in both the thick and thin disk, and broad mass distribution are also all easily consistent with our proposed scenario \citep{bergeat:2002.carbon.star.kinematics}. And the unusual isotopic ratios observed in $^{12}${\small C}$/^{13}${\small C} ($\lesssim10-20$) are almost identical in their distribution to those directly observed in presolar carbide grains (for a review, see \citealt{zinner:2014.presolar.grains.review} and The Presolar Grain Database, \citealt{hynes:2009.presolar.grain.database}) -- it will be very interesting to see if the same holds for other measured isotopic ratios in grains (such as $^{14}${\small N}$/^{15}${\small N} or $^{17,18}${\small O}$/^{16}${\small O}). Some of these observations have motivated the idea that massive grains are made in such stars; interestingly, we suggest that the causality may in fact run the oppose way. Of course, the two possibilities are not mutually exclusive; the stars probably represent the sites of new grain formation, which in turn help ``seed'' new populations, and the connection can become self-reinforcing! Moreover, alternative explanations for this sub-population, including stellar evolution processes (mixing or production from rotation, convection, unusual AGB phases or {\small He}-flashes), stellar mergers or mass transfer, all appear to make predictions in serious conflict with the observations \citep{izzard:2007.r.stars.from.binary.mergers,zamora:2009.carbon.r.star.chemistry,mocak:2009.r.star.helium.flash.sims,piersanti:2010.r.star.from.merger.sims.dont.happen}. It is much more natural to assume the stars are simply ``primordially'' enhanced in the species most abundant in large grains (if a mechanism to do so exists); the other mechanisms above would still act, of course, and modify the abundances actually measured, but they would only have to explain secondary characteristics like the {\small N} enhancement in these stars believed to result from the {\small CN} cycle and {\small He}-burning \citep[relatively easy to produce; see][]{zamora:2009.carbon.r.star.chemistry}. As noted in prediction {\bf (10)} above, the most extreme examples of the predicted populations could also appear as a subset of the observed hydrogen-deficient carbon or R Cor Bor populations.

On the other hand, while grain density fluctuations may play a role in seeding some {\small C} abundance variations in the ``N'' and ``J'' sub-populations of carbon stars, these exhibit features such as enhanced {\small Li}, s-process products, and $^{13}${\small C} abundances which are not obviously predicted by this process (and indeed, AGB mixing, mass loss, and merger process appear to explain them naturally; \citealt{abia:2000.carbon.j.star.chemistry,abia:2002.s.process.carbon.star.abundances,zhang:2013.r.star.synthesis.models}). 

In an environment with very low mean metallicity but large fluctuations in {\small [C/H]} owing to large grain density inhomogeneities, CEMPS would be especially favored to form, because the regions with enhanced dust density would be the ones that could easily cool and thus actually form stars, whereas regions with low {\small [C/H]}$\lesssim -3$ generically have cooling times much longer than their dynamical times and so cannot efficiently form stars \citep[see e.g.][and references therein]{barkana:reionization.review}. Our calculations here have generally focused on conditions closer to solar mean metallicity where cooling is always efficient, so in a future paper we will consider in more detail the specific predictions for CEMPS owing to the interplay between dust density fluctuations and cooling physics.

%\vspace{-0.5cm}
\subsection{Globular Clusters}
\label{sec:globulars}

This mechanism, occurring before the initial generation of star formation, may help to explain some of the observed star-to-star abundance variations in massive globular clusters \citep{carretta:2009.gc.abundance.var.compilation}, and resolve some tensions between the observed abundance spreads and constraints on second-generation star formation which limit quantities like the available gas reservoirs and ``donor star'' populations to the GCs \citep[see e.g.][]{muno:2006.no.gas.in.young.starclusters,sana:2010.young.starcluster.single.age,seale:2012.yso.lmc.clump.lifetime.constraints,bastian:2013.no.ongoing.sf.in.young.gcs,larsen:2012.mass.loss.self.enrichment.constraints.cluster}. 
However,  we wish to emphasize that we do not expect this scenario to provide a complete description of abundance variations in massive GCs. Enhanced {\small He}, and a tight anti-correlation between {\small Na} and {\small O}, are not obviously predicted by grain clustering. 
Since globulars are bound, massive GCs will necessarily experience other processes such as pollution by stellar mass-loss products, second-generation star formation, mergers, and complicated multiple interactions. When they occur, these sorts of enrichment processes will dominate the observed variations in these systems \citep[see e.g.][]{dercole:2008.multiple.ssp.in.gcs.from.ejecta,conroy:2011.multiple.ssp.form.in.gc,bastian:2013.early.disc.accretion.abundance.var.gc}.\footnote{That said, some globulars exhibit weak {\small Li} and {\small He} spreads \citep{piotto:2007.he.variations.gc.ngc2808,monaco:2012.lithium.variation.gc.constraints}; allowing for the fact that some of the other variations in these systems might come from the mechanism in this paper, instead of e.g.\ evolution products, would further resolve some tensions. And the predicted fluctuations from the preferential dust concentration in species like {\small Fe} are well within the observationally allowed spreads as a function of GC mass \citep{carretta:2009.iron.spread.vs.gc.mass}.}

%\vspace{-0.5cm}
\subsection{Open Clusters}
\label{sec:oc.compare}

As noted in \S~\ref{sec:predictions} point {\bf (9)} above, unlike the second-generation products invoked in GCs, we expect this process to occur identically in sufficiently massive open clusters. Measuring abundance spreads in open cluster stars is challenging and very few open clusters (especially very massive ones) have strong constraints \citep[see][]{martell:2009.gc.vs.open.abundance.var,pancino:2010.abundance.open.cluster,bragaglia:2012.berkeley.39.open.cluster.no.abundance.var,carrera:2013.open.cluster.spread.search}. In the (limited) compilations above, every OC for which significant abundance spreads can be definitively ruled out lies significantly below the threshold size/mass {\bf (1)} we predict. Interestingly, the couple marginal cases (NGC 7789, Tombaugh 2) lie near the threshold, but with large systematic uncertainties in these quantities \citep[][]{frinchaboy:2008.tombaugh2.metal.var}. 

One currently known OC (NGC 6791) definitely exhibits abundance spreads \citep[][and references therein]{carraro:2013.ngc6791.review}. Interestingly, many apparently ``anomalous'' properties of this cluster may fit naturally into our scenario. The spreads in {\small Na} and {\small O} are very large, $\sim0.7\,$dex \citep[and may be multi-modal;][]{geisler:2012.ngc6791.abundance.var}; with little spread in {\small Fe}. Particularly striking, {\small Na} and {\small O} are {\em not} anti-correlated, as they are in models where the abundance variations stem from stellar evolution products. The most metal-rich stars exhibit a highly unusual mean isotopic ratio of $^{12}${\small C}$/^{13}${\small C}$\approx10$, similar to that found in grains as in \S~\ref{sec:carbon.stars} \citep{origlia:2006.ngc6791.carbon.isotopes}. The absolute metallicity is also very high ({\small [Fe/H]$\sim0.4-0.5$}), suggesting a high dust-to-gas ratio which would make the effects described here both stronger and more directly observable (since the condensed-phase heavy element abundances should be larger). It is one of the most massive open clusters known, with mass of a few $\times10^{4}\,\msun$ and size $\gtrsim20\,$pc, making it one of the only OCs above the threshold predicted \citep{carrera:2012.ngc6791.open.cluster.abundance.var}. NGC 6791 also exhibits an anomalous blue horizontal branch stellar population; as discussed in point {\bf (10)} above, this might be explained with a sub-population of highly-enriched stars with relatively small, depleted hydrogen envelopes \citep[see e.g.][]{liebert:1994.blue.hb.stars.ngc6791,twarog:2011.ngc6791.extended.sf,carraro:2013.ngc6791.review}. If we combine the measured cluster mass and size with the PDF predicted in Fig.~\ref{fig:z.distrib} (taking $\Rgrain_\mu\approx1$ and $f_{p}=0.5$), we predict $\sim 5$ dwarf stars with $Z\gtrsim10\,Z_{\sun}$ (with large uncertainties) -- \citet{carraro:2013.ub.photometry.6791} identify $\sim 10-20$ BHB stars; an interesting coincidence.

%Following the discussion in point {\bf (10)}, we specifically plot the main sequence for pure carbon stars and perfectly-mixed $Z\grain=1$ stars from the models in \citet{paczynski:1972.carbon.stars}, as well as the main sequence for star with $Z\grain=1$ and a $\sim 10\%$ (unmixed) external H envelope from \citet{kippenhahn:2012.stellar.evolution.book} (for each case, we use the models in --- to convert the predicted $T_{\rm eff}$ and $L_{\rm bol}$ into the observed magnitudes and colors, with a distance modulus $m-M=13.65$). Unsurprisingly, given the previous inferences from these populations, the agreement is good.

%\vspace{-0.5cm}
\subsection{The UV Excess in Elliptical Galaxies}
\label{sec:uv.excess}

It is well-known that early-type galaxies exhibit an anomalous ``UV excess'' (UVX) or ``UV upturn'' -- an excess of UV light stemming from old stellar populations, generally attributed to the same rare, ``hot'' (excessively blue), metal-enhanced stars discussed in \S~\ref{sec:predictions} \&\ \S~\ref{sec:oc.compare} above \citep{horch:1992.uv.excess.origins,greggio:1999.uv.excess.review}. Such stars have been identified as abundance outliers in the metal-rich population of the Galactic thick disk \citep{thejll:1997.hot.subdwarf.thick.disk}, and in metal rich clusters like those described above \citep{liebert:1994.blue.hb.stars.ngc6791}.

In fact, the UVX appears to be empirically identical to the excess population seen in NGC 6791 -- if a couple percent of the luminosity of ellipticals were composed of stellar populations identical to NGC 6791, the integrated light would match what is observed \citep[see][]{liebert:1994.blue.hb.stars.ngc6791,landsman:1998.uv.upturn.relation.to.starclusters,brown:2004.uv.excess.evolution}. So to the extent that the model here explains the anomalies in that system, it can also account for the UVX. In contrast, it has been shown that the UVX cannot be reproduced by the sum of GC populations  \citep{van-albada:1981.uv.excess.vs.globulars,bica:1988.uv.excess.vs.starclusters,davidsen:1992.uv.excess.review,dorman:1995.uv.excess.constraints}. Most other alternative explanations have been ruled out, including: metal-poor populations \citep[references above and][]{rose:1985.ell.stellar.pop.constraints}, white dwarfs \citep{landsman:1998.uv.upturn.relation.to.starclusters}, and ``blue stragglers'' \citep{bailyn:1995.blue.stragglers.globular}. 

It is also well-established that the UVX is stronger in more metal-rich galaxies \citep{faber:1983.uv.excess.vs.linestrengths,burstein:1988.uv.excess}; this is a natural prediction of the mechanism here, since the enhanced abundances of the galaxy imply higher mean dust-to-gas ratios, making the process we describe easier and more observable (since smaller fluctuations are required to produce extreme populations, and the absolute metallicities of ``enhanced'' stars are larger). More recently, it has been observed that there is a good correlation between the UVX and {\small C} and {\small Na} abundances, while there is not a strong correlation with {\small Fe} \citep{rich:2005.uv.excess.weak.metallicity.corr}; in particular in the UVX systems it seems that the light-element abundances fluctuate {\em independent} from the heavier elements \citep[e.g.\ {\small Fe} varies much more weakly;][]{mcwilliam:1997.chemical.evol.review,worthey:1998.chemical.synthesis.ellipticals,greggio:1999.uv.excess.review,donas:2007.uv.excess.corr.ell.lenticular}, exactly as predicted here. Additional properties such as the lack of strong environmental dependence, weak redshift evolution, and large dispersion in UVX populations, are consistent with our scenario \citep{brown:2004.uv.excess.evolution,atlee:2009.uv.excess.evolution}. 

In prediction {\bf (8)}, we noted that this process may be more common at high redshift. Interestingly, there are an increasing number of suggestions of anomalous ``hot'' stellar populations in high-redshift galaxies. In particular, \citet{steidel:2014.hard.ionizing.field.old.stars.must.be.hot} argue that fitting the spectra of star forming galaxies at redshifts $z\gtrsim2$ seems to require a population of old ($>10\,$Myr) stars which can still have effective temperatures $T\gtrsim 30,000$\,K, whose fractional luminosity (relative to the galaxy total) increases at higher redshifts (at fixed galaxy mass) -- whatever these stars are, they are not present in ``standard,'' single-metallicity stellar population models.

\subsection{Galaxy-Wide Abundance Variations}
\label{sec:galaxy.abundance.var}

What are the consequences of this mechanism for galaxy-wide stellar abundance variations? If we assume that all GMCs lie on the mean Larson's law relations fitted in \citet{bolatto:2008.gmc.properties}, and adopt the nearly-universal Milky Way GMC mass function determination in \citet{blitz:2004.gmc.mf.universal} as generic, then $\approx 10-20\%$ of the mass in MW GMCs is in those large enough meet our criterion {\bf (1)} (for $a_{\mu}\approx 0.1$ and $T\sim 30\,$K) and produce ``strong'' fluctuations (akin to the $R_{10} > 1$ cases in Fig.~\ref{fig:z.distrib}) following Eq.~\ref{eqn:S.zgrain} with up to $\sim 1\,$dex scatter ($S_{\ln\,Z\grain}\approx 4.9$) in $Z\grain/\langle Z\grain \rangle$. We emphasize that since the GMC mass function scales as $d{N}/dM\propto M^{-1.8}$, this is a very small fraction of the total GMC population {\em by number}. Assuming a constant star formation efficiency per cloud, this translates into the same fraction of new stars formed under the relevant conditions. If we further assume that $\sim 1/3-1/2$ the heavy elements are in large grains, then this predicts a variance $S_{\ln{Z}} \sim (0.1-0.2)\times (0.3-0.5)^{2} \times (4.9/2)$ (the last factor $=2$ comes because the negative half of the density fluctuations in Fig.~\ref{fig:z.distrib} are suppressed by the abundance of gas-phase metals), or a $1\,\sigma$ dispersion of $\sim 0.05-0.1$\,dex. If $a_{\mu}\sim1$, then $\sim 40\%$ of the GMC mass could be involved, leading to $\sim 0.1-0.2$\,dex scatter. 

In comparison, in the MW and Andromeda, the dispersion in {\small Fe} (which is more weakly effected by the mechanism we propose here than the light elements) is $\sim 0.1-0.2$\,dex at fixed location in the disk and stellar population age \citep{nordstrom:2004.geneva.solar.neighborhood.stellar.survey,holmberg:2007.geneva.solar.neighborhood.survey,casagrande:2011.geneva.survey.update.metal.rich.wing.of.local.stars,duran:2013.local.age.metallicity.relation,nidever:2014.mw.metallicity.gradient.study,bensby:2014.mw.metallicity.gradient.study}, and this can rise to $\sim 0.3-0.4$\,dex in other galaxies \citep{koch:2006.mdf.carina.dwarf}. This is reassuringly consistent with our estimate, when we consider that there are other sources of scatter in abundances such as inhomogeneous mixing. And the scatter appears to grow in the older stellar populations (to $\sim 0.3-0.5\,$dex for ages $\sim 10\,$Gyr; \citealt{bensby:2013.microlensed.star.metallicity.distribution.high.metals}), perhaps evidence of larger GMCs in the early, gas-rich history of the Milky Way, which would enhance this process per our prediction {\bf (8)}. The total abundance variation observed across the Galaxy is much larger, owing to radial gradients (which contribute another $\gtrsim 0.3$\,dex scatter galaxy-wide; see \citealt{mehlert:ssp.gradients,reda:ssp.gradients,sanchezblazquez:ssp.gradients}). The scatter in light-element abundances appears to be larger, $\sim 0.2-0.3\,$dex at fixed radius and age in oxygen, for example \citep{reddy:2006.thick.disk.abundances,korotin:2014.oxygen.in.mw.vs.radius.large.scatter}, consistent with our prediction {\bf (4)}. 

Just within $\sim 150\,$pc of the sun, for example, $\sim 1$ per $1000$ stars can reach metallicities in {\small [Fe/H]}, {\small [C/H]}, or {\small [O/H]}$\sim 0.4-0.6$ \citep{haywood:2001.solar.neighborhood.metallicity.distribution,ibukiyama:2002.disk.star.local.metal.distrib,pompeia:2003.bulgelike.dwarf.star.abundance.correlations,mcwilliam:2008.oxygen.magnesium.scatter,duran:2013.local.age.metallicity.relation,hinkel:2014.hypatia.abundances.catalog}; of course, this is exactly what one expects for $\sim 0.15\,$dex scatter, and about that same scatter is seen in the {\em ratios} of the light elements ({\small C, O}) to {\small Fe}. Many stars with metallicities {\small [Z/H]}$\gtrsim 0.4-0.5$ have now been confirmed \citep[and many are old, meaning that they deviate from the age-metallicity relation by up to $\sim 1$\,dex; see][]{feltzing:2001.super.metal.rich.stars,taylor:2002.supermetallicity.tests,taylor:2006.supermetallicity.star.tests,chen:2003.local.metal.rich.stars,carretta:2007.ngc6791.abundances}. Meanwhile non-standard selection methods strongly suggest that higher-metallicity stars exist but have been either overlooked or removed owing to strong selection effects in traditional stellar metallicity studies \citep[see][]{haywood:2001.solar.neighborhood.metallicity.distribution,cohen:2008.microlensing.bulge.star.abundances,bensby:2013.microlensed.star.metallicity.distribution.high.metals}.\footnote{In fact  extreme examples of stars with {\small [C/H]}$\gg1$ and {\small [C/Fe]}$\gg1$ are known, but these are usually dismissed from abundance surveys as ``peculiar'' products of mergers, ``re-ignited'' white dwarfs, or other special circumstances \citep[see e.g.][]{kameswara-rao:2007.rcorbor.abundance.review}.}

What about the most extreme examples (the ``totally metal'' stars)? The massive versions ($\gtrsim 2-3\,M_{\sun}$) of such objects would be unobservable. If $\sim 10\%$ of GMC mass is in clouds which can produce very strong dust clustering, and (based on our estimates in \S~\ref{sec:fluct.pdf}) one in $\sim 10^{3}-10^{4}$ cores in such clouds can reach $Z\grain \sim 1$, and we further assume a normal \citet{chabrier:imf} stellar IMF (and Galactic star formation rate $\sim 1\,\msun\,{\rm yr^{-1}}$) and (crudely) estimate the lifetimes of these stars to be similar to the theoretical helium main sequence in \citet{kozowski:1973.helium.carbon.stellar.mixes} (see point {\bf (10)}), we expect $\sim1$ such star presently in the entire Galaxy! Compare this to $\sim 10^{6}$ O-stars. However, for dwarfs with $\sim 0.3-2\,\msun$, the same estimates above, with a $\sim 10^{9}\,$yr hydrogen-burning phase, imply one in $\sim 10^{6}$ stars in the Milky Way could be a member of this sub-population. Unfortunately, existing samples in large, unbiased metallicity surveys are too small to identify such rare populations; however, this abundance is perfectly consistent with the extrapolation of a log-normal fit to the observed metallicity distribution.

\begin{figure}
    \centering
    \plotonesize{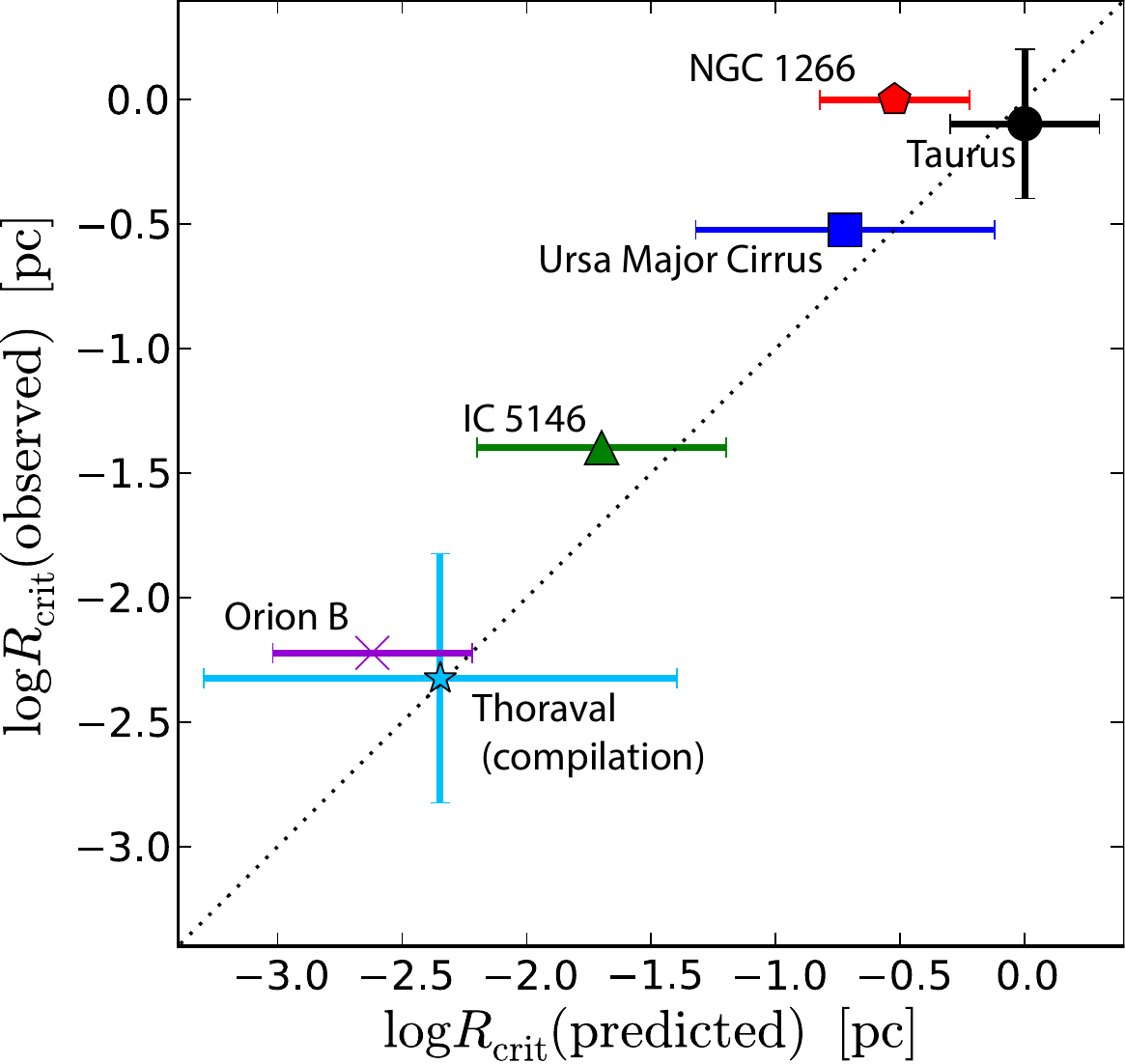}{0.95}
    %\vspace{-0.2cm}
    \caption{Observed versus predicted local fluctuations in grain abundances in nearby molecular clouds (see \S~\ref{sec:dust.to.gas.in.gmcs}). We compile observed clouds where large (factor $\sim2-5$) fluctuations in the local abundance of grains relative to gas are measured; for each, we compare the maximum observed scale of the fluctuations ($R_{\rm crit}$(observed)) to the characteristic scale $R_{\rm crit}$(predicted) of fluctuations we would predict from turbulent concentration (Eq.~\ref{eqn:rcrit.vs.cloudparameters}), given the cloud properties. Dotted line shows perfect agreement. For Taurus, we take properties ($R_{10}\approx 3$, $M_{\rm cl}\approx 2\times10^{4}\,\msun$, $\Rgrain_\mu=0.1$, $Q\sim1$, $T_{10}\sim1$) from \citet{pineda:2010.taurus.large.extinction.variations}. NGC 1266 ($T_{10}\sim 10$, $\Sigma_{300}\sim 40$, $Q\sim 2$, $R_{10}\sim 10$, $\Rgrain_\mu\sim1$) observations come from  \citet{pellegrini:2013.ngc.1266.shocked.molecules} and \citet{nyland:2013.radio.core.ngc1266}. Orion B ($M_{\rm cl}\sim10^{5}\,\msun$, $\pi\,R_{\rm cl}^{2}=19$\,deg$^{2}$ at $400\,$pc, so $R_{10}=3$, $\Sigma_{300}=0.114$, $T_{10}\sim1$, $\Rgrain\sim10\,$nm) data is from \citet{abergel:2002.size.segregation.effects.seen.in.orion.small.dust.abundances}. \citet{miville-deschenes:2002.large.fluct.in.small.grain.abundances} measure the Ursa Major cirrus ($\Rgrain_\mu\sim0.1$, $n=3\,M_{\rm cl}/(\mu\,m_{p}\,4\pi\,R_{\rm cl}^{3}) \sim 100\,{\rm cm^{-3}}$; $T\sim200\,$K, with turbulent $\Delta v\sim5-10\,{\rm km\,s^{-1}}$ at scales $R=10\,$pc; we calculate directly from this the scale where $\tstop\approx \Teddy(R_{\rm crit})$). IC 5146 ($n\sim500\,{\rm cm^{-3}}$, $T_{10}\sim1$, $\Delta v\approx 1\,{\rm km\,s^{-1}}$ on scales $\approx 0.06\,$pc, $\Rgrain_\mu\sim0.1$) from \citet{thoraval:1997.sub.0pt04pc.no.cloud.extinction.fluct.but.are.on.larger.scales}. The point labeled ``Thoraval'' is a compilation of sightlines through CGCG525-46, IR04139+2737, and G0858+723 from \citet{thoraval:1999.small.scale.dust.to.gas.density.fluctuations}, with $n\sim100-1000\,{\rm cm^{-3}}$, $\Rgrain_\mu\sim0.1-1$, $T_{10}\approx1$, and rms $\Delta v\sim0.1\,{\rm km\,s^{-1}}$ across $0.01\,$pc. The ``predicted'' error bars show the allowed range corresponding to the range in quoted cloud parameters. Observed error bars are shown where given by the authors. All of these values have large systematic uncertainties; however, the agreement is encouraging. 
    \label{fig:observed.fluctuations}}
\end{figure}

%\vspace{-0.5cm}
\subsection{Dust-to-Gas Variations in GMCs and Nearby Galaxies}
\label{sec:dust.to.gas.in.gmcs}

The model here predicts large variations in the dust-to-gas ratio in the star-forming molecular gas, under the right conditions. But as noted in {\bf (3)}, these fluctuations will not necessarily manifest in the most obvious manner (via single-wavelength extinction mapping). However, there are some tracers sensitive to large grains, for example in the sub-millimeter. And there are a number of direct hints of this process ``in action.'' It is well known that clear variations of the extinction curve shape and sub-mm spectral index -- indicating a different relative abundance of small and large grains -- are observed across different sightlines in the Milky Way, LMC, and SMC \citep{weingartner:2001.dust.size.distrib,gordon:2003.large.variations.extinction.curves.in.lmc.smc.mw.sightlines}. \citet{dobashi:2008.dust.gas.variations.observed.over.lmc.gmcs} and \citet{paradis:2009.large.variation.across.lmc.in.dust.size.distrib} specifically show how this implies substantial variations in the size distribution in different large LMC clouds. 

Measuring this {\em within} clouds is more challenging, but has been done. \citet{thoraval:1997.sub.0pt04pc.no.cloud.extinction.fluct.but.are.on.larger.scales,thoraval:1999.small.scale.dust.to.gas.density.fluctuations} look at fluctuations of the abundance of smaller grains ($\Rgrain_\mu\sim0.1$) within several nearby clouds (IC 5146,CGCG525-46, IR04139+2737, G0858+723), and find significant (factor $>2$) fluctuations in the local dust-to-gas ratio within the clouds on scales $\sim 0.001-0.04\,$pc. They specifically noted that these size scales corresponded, for the cloud properties and grain sizes they examined, to the ``resonant'' scale for marginal turbulent coupling (our $R_{\rm crit}$). Similar effects have been observed in other clouds. With more detailed models of the grain sizes, \citet{miville-deschenes:2002.large.fluct.in.small.grain.abundances} found factor $\sim 5$ fluctuations in the ratio of small-to-large grains on $\sim 0.1\,$pc scales in the Ursa Major cirrus. \citet{flagey:2009.taurus.large.small.to.large.dust.abundance.variations} and \citet{pineda:2010.taurus.large.extinction.variations} see similar (factor $\sim 5$) variations in the small-to-large grain abundance ratio across scales from $\sim 0.1-1\,$pc in the extended Taurus cloud (from full modeling of the grain size distribution along each sightline, as well as extinction-to-CO variations). This region is $\sim 30\,$pc across, with mean surface density $\sim 10\,M_{\sun}\,{\rm pc^{-2}}$, so our predicted $R_{\rm crit}$ also reaches $\sim 1\,$pc -- given its proximity and well-studied nature, Taurus may be an ideal laboratory to test this process. \citet{abergel:2002.size.segregation.effects.seen.in.orion.small.dust.abundances} map fluctuations in the abundance of very small particles ($\sim 10\,$nm) in Orion B; they see large fluctuations, but (as predicted) on smaller scales ($\sim 0.006$\,pc). Alternative interpretations of these observations, such as differential (location-dependent) icing and coagulation or shattering, have also been discussed; nonetheless the scales appear strikingly similar to the predictions of turbulent concentration. Larger grains ($\Rgrain_\mu\sim1$) may have been observed to fluctuate in abundance in the extreme nuclear region of NGC 1266. Several authors \citep[see e.g.][]{alatalo:2011.ngc1266.molecular.outflow,pellegrini:2013.ngc.1266.shocked.molecules,nyland:2013.radio.core.ngc1266} have noted that the dust traced in sub-millimeter imaging and nuclear molecular gas are not necessarily co-spatial on $\sim 1\,$pc scales (although there are other potential explanations, such as shocks); the properties of the region imply, according to our calculations, large fluctuations in sub-clouds with sizes above a few parsec (masses $\gtrsim10^{6}\,\msun$). 

In Fig.~\ref{fig:observed.fluctuations}, we summarize these observations, in each case using the parameters of the cloud given by the authors to estimate our theoretically predicted $R_{\rm crit}$ for the grain sizes measured. The uncertainties are large; however, the agreement over several orders of magnitude in cloud size is encouraging.

\subsection{Consequences for Planet Formation}

If this mechanism operates, it can have dramatic implications for planet formation in the proto-stellar/proto-planetary disks which form from the affected cores. Many models of terrestrial planet formation, for example, predict far more efficient planet formation at enhanced metallicities (via either simplistic grain growth/accretion, or the onset of instabilities such as the streaming instability; see references in \S~\ref{sec:intro}). For example, in the simulations of \citep{bai:2010.grain.streaming.vs.diskparams}, and analytic models in e.g.\ \citep{cuzzi:2010.planetesimal.masses.from.turbulent.concentration.model,hopkins:2014.pebble.pile.formation}, there is an exponential increase in the maximum density of solids reached (enough to allow some to become self-gravitating planetesimals) if the initial disk metallicity is a few times solar -- this is sufficient such that, once the metals grow in the disk into more massive grains and sediment into the midplane, their density becomes larger than that of the midplane gas, triggering a range of new dust-clumping instabilities. Such metallicity enhancements are relatively modest, compared to the extremes we have discussed above. And in the most extreme cases we predict ($Z\sim 100\,Z_{\sun}$), these instabilities would operate not just in the midplane but everywhere in the disk, almost instantaneously upon its formation! 

Even if the mechanisms of planet formation remain uncertain, there is an increasingly well-established correlation between giant planet occurrence rates and stellar metallicity \citep[see e.g.][and references therein]{johnson:2010.giant.planet.rate.vs.metallicity}. Whatever causes this, it is clear that stellar abundance variations -- seeded before both the star and protoplanetary disk form -- must be accounted for in understanding the occurrence rates and formation conditions of giant planets. One might predict, for example, a dramatic increase in the occurrence rates of planets in the sorts of open clusters which formed in very massive, large clouds, similar to NGC 6791 discussed above; in these circumstances the occurrence rate might have more to do with the statistics of seeded dust abundance variations, than with the planet formation mechanisms themselves.

Particularly interesting, our model does not just predict a uniform variation in metallicities (as is usually assumed when modeling planet formation under ``high'' or ``low'' metallicity conditions). Rather, we specifically predict abundance variations in the species preferentially concentrated in the largest grains in large, cold clouds. So we predict there should be disks preferentially concentrated in carbonaceous grains, for example. Similarly, if ice mantles form, there will be disks with over-abundances of oxygen. And if the progenitor cloud was slightly larger still, silicates can be preferentially enhanced (relative to species like iron). So it is actually possible, under the right conditions, to form a proto-planetary disk which is highly enhanced in large carbonaceous and silicate grains, even while the host star exhibits apparently ``normal'' abundances of iron and most other species. This will radically change the chemistry of the massive grains which form in the disk, hence the conditions (and mechanisms) of planet formation, as well as the composition of the resulting planets!

%\vspace{-0.5cm}
\section{Future Work}

Our intent here is to highlight a new and (thus far) unexplored physical process by which unusual abundance patterns may be seeded in stars formed in massive molecular clouds. Many consequences of this should be explored in more detail.

Predicting more accurately the abundance patterns imprinted by this process requires combining the calculations above (for grain density fluctuations as a function of grain size) with detailed, explicit models for the grain chemistry. Specifically, knowing the abundance of different elements, integrated over all large grains above some minimum $\Rgrain_\mu$, would enable strong quantitative predictions for relative fluctuation amplitudes of different abundances. At the moment this is extremely uncertain and model-dependent, since the chemical structures of large grains in particular are difficult to probe and often degenerate (for a discussion, see \citealt{draine:2003.dust.review,zubko:2004.dust.chemistry.models}). 

Another critical next step is to directly simulate the formation of proto-stellar cores in a GMC while explicitly following a size distribution of grains as aerodynamic particles. This would remove many current uncertainties in the non-linear grain clustering amplitudes: the analytic model here is a reasonable approximation to existing simulations but cannot capture the full range of behaviors and subtle correlations between the velocity and density fields in turbulence, as well as the more complicated mutual role of gas density fluctuations seeding core formation while independent grain density fluctuations occur alongside. Probably the most poorly-understood element of the physics here is how the instabilities we describe are modified in the presence of a magnetic field (\S~\ref{sec:preconditions}). If most of the grains are weakly charged (which is by no means certain); we would then expect them to go through alternating phases of neutral and charged as they collide with electrons and ions, ``seeing'' a non-linearly fluctuating magnetic force (while a core collapses through fields via ambipolar diffusion). Properly modeling this requires (in addition to the physics above)  magneto-hydrodynamic turbulence with non-ideal MHD (given the ionization fractions in regions of interest), explicit treatment of grain-electron interactions/capture, and subsequent capture-dependent grain-MHD interactions. 

But any such simulations remain very challenging. Almost all current molecular cloud simulations treat grains (if at all) by assuming either a constant dust-to-metals ratio, or as a fluid (the two-fluid approximation; see e.g.\ \citealt{downes:2012.multifluid.turb.density.pdf}). In these approximations, the physical processes driving dust-to-gas ratio fluctuations (the subject of this paper) are artificially prohibited. It has only just become possible to follow grains as aerodynamic species in astrophysical codes (the most basic requirement to see the behavior here), and while this has been applied to idealized fluid dynamics and proto-planetary disk simulations (\S~\ref{sec:intro}) this has not yet been extended to include all the processes relevant in GMCs. For example, no such simulations have included all the MHD effects described above. The resolution also remains challenging; simulations must be able to resolve sub/trans-sonic turbulent eddies with scales down to about $\sim0.05$ times the critical scale ($\Rcrit$) for convergence \citep{hopkins:2013.grain.clustering}, which here is the scale of individual cores, while still capturing the cloud-scale dynamics such that a statistical population of cores can be tracked. However, some early numerical experiments of turbulent boxes with ideal MHD and simplified grain-gas coupling in \citet{lazarian:2004.mhd.effects.dont.dramatically.change.turb.concentration} argue that MHD effects do not significantly alter grain concentration, since motion perpendicular to field lines is still dominated by vorticity effects. But \citet{decamp:2002.turb.concentration.ism.conditions.chemistry.fx.make.stronger} also argue that including non-equilibrium grain chemistry can actually enhance grain clustering under typical cloud conditions. So more work is clearly needed.

It is also important to explore the predictions for stellar properties. Stellar evolution models have not, in general, considered the case of stars forming from regions super-enhanced in certain species, let alone stars with an order-unity {\em initial} mass fraction in heavy elements. Knowing whether these would simply appear as stars which have completed light-element burning (and thus may already be in some observed samples, but un-recognized), or would produce unique observational signatures, would enable powerful tests of the scenario outlined here as well as providing a new window into extreme stellar physics.

%\vspace{-0.7cm}
\acknowledgments 
We thank Charlie Conroy, Selma de Mink, and Jessie Christiansen for many helpful discussions and sanity checks during the development of this work. We also thank Todd Thompson, Chris Matzner, Eric Pellegrini, Norm Murray, and Nick Scoville for some detailed discussions of the physical consequences for star formation and suggestions for observational comparisons and tests. We thank Matt Kunz, Avishai Gal-Yam, and Steve Longmore for helpful comments, including several ideas for follow-up work, and Peter Goldreich for stimulating discussions on the general physics involved. And we thank Alvio Renzini for highlighting some unclear statements that may have been potential points of confusion. Partial support for PFH was provided by the Gordon and Betty Moore Foundation through Grant \#776 to the Caltech Moore Center for Theoretical Cosmology and Physics. \\

%\vspace{-0.1cm}
\bibliography{/Users/phopkins/Documents/work/papers/ms}

\begin{appendix}

\section{Scalings in the Super-Sonic Limit}

The scalings we used in the main text to derive the grain stopping time $\tstop$ assumed the grain is not moving super-sonically {\em relative to the gas atoms with which it collides}; similarly, the model for the grain clustering statistics was derived assuming the local flow moving with the grain has relative motions which are sub or trans-sonic. This was done for good reason: because the scale of clustering we are interested in is the sonic length -- the scale where the turbulence becomes sub-sonic -- this is the most appropriate. However, we can re-derive the appropriate scalings assuming {\em all} scales behave as super-sonic turbulence, and obtain identical results. We show this here. 

In the limit where the grains are moving highly super-sonically relative to the gas with which they collide, the stopping time is modified to become
\begin{align}
\label{eqn:tstop.supersonic} \tstop \rightarrow \tstop^{SS} = \frac{\rhointernal\,\Rgrain}{\rhogas\,\Delta v_{\rm gas-grain}}
\end{align}
where $\Delta v_{\rm gas-grain}$ is the rms relative velocity of the dust and gas as we take the separation/averaging scale around the grains $\rightarrow 0$ \citep{draine:1987.grain.charging}. For grains in a turbulent medium, this is giving by integrating over all modes/eddies of different sizes, accounting for the partial coupling of grains to gas, and has been calculated (and simulated) by many authors \citep[see][]{voelk:1980.grain.relative.velocity.calc,markiewicz:1991.grain.relative.velocity.calc,ormel:2007.closed.form.grain.rel.velocities,pan:2013.grain.relative.velocity.calc}. At the level of accuracy we require here, these calculations all give approximately $\Delta v_{\rm gas-grain} \sim \Veddy(\tstop\sim \Teddy)$, i.e.\ the grains are accelerated to relative motion versus the gas corresponding to the ``resonant'' eddies (since larger eddies simply entrain grains, so they move with gas with no relative motion, and smaller eddies do not significantly perturb the grains). The problem is, $\Delta v_{\rm gas-grain}$ depends on $\tstop$, which depends on $\Delta v_{\rm gas-grain}$. However, this simply turns Eq.~\ref{eqn:tstop.supersonic} into a non-linear equation which is easily solved. Since, for super-sonic turbulence, $\Veddy \sim v_{\rm cl}\,(R/R_{\rm cl})^{1/2}$ and $\Teddy \sim t_{\rm cl}\,(R/R_{\rm cl})^{1/2}$ (where $t_{\rm cl}\equiv R_{\rm cl}/v_{\rm cl}$), we have $\Veddy(\tstop\sim \Teddy) \sim (\tstop/t_{\rm cl})\,v_{\rm cl}$. Plugging this into Eq.~\ref{eqn:tstop.supersonic}, we obtain 
\begin{align}
\tstop^{SS} &= \frac{\rhointernal\,\Rgrain}{\rhogas\,\Delta v_{\rm gas-grain}} \sim \frac{\rhointernal\,\Rgrain}{\rhogas\,v_{\rm cl}\,(\tstop^{SS}/t_{\rm cl})} \\ 
\tstop^{SS} &\sim \left( \frac{\rhointernal\,\Rgrain\,t_{\rm cl}}{\rhogas\,v_{\rm cl}} \right)^{1/2} = t_{\rm cl}\,\left[\left(\frac{\rhointernal\,\Rgrain}{\rhogas\,\cs}\right)\,\frac{\cs}{t_{\rm cl}\,v_{\rm cl}} \right]^{1/2} = t_{\rm cl}\,\left( \frac{\tstop(\Delta v=\cs)}{t_{\rm cl}}\,\frac{\cs}{v_{\rm cl}}  \right)^{1/2}
\end{align}

We can now use this to revisit the key equations in the text. The ratio of stopping time to eddy turnover time becomes
\begin{align}
{\Bigl\langle}\frac{\tstop^{SS}}{t\eddy(R)}{\Bigr\rangle} &\approx 0.1\,\left(\frac{\Rgrain_\mu}{\Sigma_{300}} \right)^{1/2}\,{\Bigl(} \frac{R}{R_{\rm cl}} {\Bigr)}^{-1/2}
\end{align}
Note that this scales in the same manner with $(R/R_{\rm cl})$ as in the text, but with a factor $\approx 2$ different pre-factor. 
The critical scale for clustering is still the scale where $\tstop^{SS}\sim \Teddy$, so we set this to unity and obtain 
\begin{align}
\Rcrit^{SS} \sim 0.1\,{\rm pc}\,\frac{\Rgrain_{\mu}\,R_{10}}{\Sigma_{300}} \approx 0.1\,{\rm pc}\,\Rgrain_{\mu}\,\left(\frac{400\,{\rm cm^{-3}}}{\langle n_{\rm cl} \rangle}\right)
\end{align}
Interestingly, in this limit, the critical scale for clustering is similar to that which we derived in the main text, but depends only on the grain size and three-dimensional density of the cloud (the temperature and $Q$ dependence of the cloud disappear because the cloud temperature is not important). For clouds with low mean densities, $\langle n_{\rm cl} \rangle\sim 10$, the clustering scale can reach $\sim 4\,$pc, while for super-dense clouds, we can have $R_{\rm crit} \ll 0.1\,$pc. 

In a supersonic cloud, the sonic length is still given by the same scaling ($R_{s}\approx 0.013\,{\rm pc}\,T_{10}/(Q^{2}\,\Sigma_{300})$), and this still determines the characteristic size where gas density fluctuations cease and protostellar cores form \citep[see][]{hopkins:frag.theory}. So we calculate the ratio of the grain clustering scale to sonic length and obtain 
\begin{align}
\frac{R_{\rm crit}^{SS}}{R_{s}} \approx 4.8\,\frac{\Rgrain_{\mu}\,R_{10}\,Q^{2}}{T_{10}} = \left(\frac{R_{\rm crit}}{R_{s}}\right)^{1/2}_{\Delta v_{\rm gas-grain} = \cs}
\end{align}

The critical cloud size above which $R_{\rm crit}^{SS} > R_{s}$, then, is given by solving this to obtain: 
\begin{align}
R_{\rm cl} &\gtrsim \frac{2}{3\pi\,Q^{2}}\,\frac{\cs^{2}}{G\,\rhointernal\,\Rgrain}\\
R_{10} &\gtrsim 0.2\,\Rgrain_{\mu}^{-1}\,Q^{-2}\,T_{10}
\end{align}
This is {\em exactly} the same as we obtained in the text! The reason is simple: what we solve for is the cloud where the grain clustering scale equals the sonic scale, i.e.\ the scale where $\cs = v_{t}(R)$. When $v_{t}(R)=\cs$, though, the ``super-sonic'' and ``sub-sonic'' stopping times ($\tstop^{SS}$ and $\tstop$) are identical -- as they must be. 

Furthermore, as discussed in the text, the generation of grain clustering under these conditions does not depend on whether the gas motion is super or sub-sonic, only that there is a non-zero vorticity field, and ``resonant'' structures exist with vorticity $|\nabla \times {\bf v}_{\rm gas}| \sim \tstop^{-1}$. In super-sonic turbulence, although the global {geometric} structure of the flow may appear different from sub-sonic turbulence, local solenoidal structures are constantly formed by shear motions of gas, and in fact they contain $\sim 2/3$ of the power \citep[as opposed to all of it, in sub-sonic turbulence; see][]{federrath:2008.density.pdf.vs.forcingtype,konstantin:mach.compressive.relation}. To first approximation, the detailed geometry of these structures is not important -- in fact, in the analytic models used to derive the estimated variance in the dust-to-gas-ratio induced by such structures, it is assumed to be random (more appropriate, in fact, in the super-sonic case).

\end{appendix}

\end{document}